\documentclass[twocolumn]{aastex631}

\received{April 3, 2023}
\revised{May 9, 2023}
\accepted{May 9, 2023}

\submitjournal{ApJ}

\shorttitle{The Merging of a Coronal Dimming and a Coronal Hole}
\shortauthors{Ngampoopun et al.}

\graphicspath{{./}{Figure/}}
\usepackage{booktabs}
\usepackage{multirow}
\usepackage{graphicx}

\begin{document}
\title{The Merging of a Coronal Dimming and the Southern Polar Coronal Hole} 

\correspondingauthor{Nawin Ngampoopun}
\email{nawin.ngampoopun.21@ucl.ac.uk}

\author[0000-0002-1794-1427]{Nawin Ngampoopun}
\affiliation{University College London, Mullard Space Science Laboratory, Holmbury St. Mary, Dorking, Surrey, RH5 6NT, UK}

\author[0000-0003-3137-0277]{David M. Long}
\affiliation{Astrophysics Research Centre, Queen’s University Belfast,
University Road, Belfast, Northern Ireland, BT7 1NN, UK}
\affiliation{University College London, Mullard Space Science Laboratory, Holmbury St. Mary, Dorking, Surrey, RH5 6NT, UK}

\author[0000-0002-0665-2355]{Deborah Baker}
\affiliation{University College London, Mullard Space Science Laboratory, Holmbury St. Mary, Dorking, Surrey, RH5 6NT, UK}

\author[0000-0002-0053-4876]{Lucie M. Green}
\affiliation{University College London, Mullard Space Science Laboratory, Holmbury St. Mary, Dorking, Surrey, RH5 6NT, UK}

\author[0000-0003-2802-4381]{Stephanie L. Yardley}
\affiliation{Department of Meteorology, University of Reading, Reading, UK}
\affiliation{University College London, Mullard Space Science Laboratory, Holmbury St. Mary, Dorking, Surrey, RH5 6NT, UK}
\affiliation{Donostia International Physics Center (DIPC), Paseo Manuel de Lardizabal 4, 20018, San Sebastián, Spain}

\author[0000-0001-7927-9291]{Alexander W. James}
\affiliation{University College London, Mullard Space Science Laboratory, Holmbury St. Mary, Dorking, Surrey, RH5 6NT, UK}
\affiliation{European Space Agency, European Space Astronomy Centre, Spain}

\author[0000-0003-0774-9084]{Andy S.H. To}
\affiliation{University College London, Mullard Space Science Laboratory, Holmbury St. Mary, Dorking, Surrey, RH5 6NT, UK}

\begin{abstract}
We report on the merging between the southern polar coronal hole and an adjacent coronal dimming induced by a coronal mass ejection on 2022 March 18, resulting in the merged region persisting for at least 72 hrs.  We use remote sensing data from multiple co-observing spacecraft to understand the physical processes during this merging event. The evolution of the merger is examined using Extreme-UltraViolet (EUV) images obtained from the Atmospheric Imaging Assembly onboard the \textit{Solar Dynamic Observatory} and Extreme Ultraviolet Imager onboard the \textit{Solar Orbiter} spacecraft. The plasma dynamics are quantified using spectroscopic data obtained from the EUV Imaging Spectrometer onboard \textit{Hinode}. The photospheric magnetograms from the Helioseismic and Magnetic Imager are used to derive magnetic field properties. To our knowledge, this work is the first spectroscopical analysis of the merging of two open-field structures. We find that the coronal hole and the coronal dimming become indistinguishable after the merging. The upflow speeds inside the coronal dimming become more similar to that of a coronal hole, with a mixture of plasma upflows and downflows observable after the merging. The brightening of bright points and the appearance of coronal jets inside the merged region further imply ongoing reconnection processes. We propose that component reconnection between the coronal hole and coronal dimming fields plays an important role during this merging event, as the footpoint switching resulting from the reconnection allows the coronal dimming to intrude onto the boundary of the southern polar coronal hole.

\end{abstract}

\section{Introduction} \label{sec:intro}
Coronal holes (CHs) are regions of relatively low density and temperature in the solar corona, appearing dark when observed in Extreme-Ultraviolet (EUV) and soft X-ray passbands \citep[see review by ][and references therein]{Cranmer2009}. Their main characteristic is the open magnetic field configuration, which allows the plasma to escape the corona as the solar wind \citep{Altschuler1977,Levine1977}. CHs generally have lifetimes of up to several solar rotations \citep{Harvey2002, Heinemann2018b}, and they can be found in both polar regions (polar CH) and low-latitude regions (equatorial CH). It is thought that polar CHs are formed gradually due to the result of open flux accumulation and transport throughout the solar cycle, usually from the poleward migration of open fields from lower latitude \citep{Harvey2002, Lowder2017}. In contrast, the formation of isolated, low-latitude CHs was observed to be related to active region evolution during either the emergence phase \citep{Harvey1998, Wang2010} or the decaying phase \citep{Karachik2010}. CHs have been widely accepted as the source of the fast solar wind, and CH boundaries have been proposed as one of the sources of slow solar wind \citep[see, e.g.,][and references therein]{Abbo2016, Cranmer2017}. The fast solar wind streams are thought to originate from open magnetic flux tubes concentrated at the supergranular network boundaries, where they show the outflow velocities of $\sim$ 10 km s$^{-1}$ as determined from spectroscopic observation using emission lines forming at the base of the corona \citep{Hassler1999, Tu2005}. The slow solar wind from CH boundaries, on the other hand, is thought to be associated with the highly expanded open field near the CH edges \citep{Wang1990}, or be driven by interchange reconnection at open-closed magnetic field boundaries between CH and the surrounding regions \citep{Antiochos2011, Abbo2016, Aslanyan2022}. Interchange reconnection also affects the evolution of CH areas \citep{Baker2007, Kong2018} and helps maintain the quasi-rigid rotation of CHs \citep{Wang2004,Yang2011}. Coronal jets, the small, transient collimated plasma ejections frequently observed inside CHs, may also play an important role in supplying mass and energy to the solar wind\citep{Raouafi2023}.

Some dark regions in the solar corona resemble coronal holes, but they have a more transient nature and shorter lifespans. These features are historically called transient coronal holes \citep{Rust1983} but now are more commonly called coronal dimmings \citep{Hudson1996}. They can be observed as regions of reduced emission in soft X-ray and EUV passbands similar to CHs. 
Their appearance is usually related to solar eruptions, such as coronal mass ejections (CMEs) or solar flares \citep{Bewsher2008, Dissauer2018a, Dissauer2019}. Coronal dimmings can be categorised into core and secondary dimmings \citep{Mandrini2007}. Core dimmings have substantially reduced emissions and often appear in pairs localised close to the eruption site. They are interpreted as the footpoints of erupting flux ropes. Secondary dimmings, in contrast, are more widespread and have a shorter lifespan and weaker intensity reductions. They are understood to be the result of plasma escaping along the expanding overlying field arcades above the flux ropes.

It is believed that the formation of coronal dimmings is due to density depletion as the plasma escapes from the solar corona along the expanding CME structure \citep{Hudson1996,Thompson2000}. This is supported by the spectroscopic observation of strong plasma outflows of up to a hundred km s$^{-1}$ \citep[e.g.,][]{Harrison2000, Tian2012, Veronig2019}. Differential emission measure analysis (DEM) of coronal dimmings shows significant plasma density decreases of up to 50-70 \% \citep{Cheng2012, Vanninathan2018}. It is also thought that coronal dimmings contain a quasi-open magnetic field structure due to CMEs expansion, and the upflows may then contribute to the solar wind \citep{Harra2007, Attrill2010a, Lorincik2021}. 

Interchange reconnection is also proposed to play an essential role in the evolution of coronal dimmings \citep{Attrill2006, Krista2013} and their recovery after eruption \citep{Attrill2008}. \citet{Liu2007} and \citet{Gutierrez2018} observed the formation of CHs from long-lived coronal dimmings, suggesting that CHs can also form through an abrupt opening of the magnetic field by solar eruptions \citep{Heinemann2018b}.

In this paper, we report a rare event of the merging between a southern polar CH and the coronal dimming resulting from a coronal mass ejection on 2022 March 18. This event occurred during the first perihelion in the science phase of the ESA/NASA's \textit{Solar Orbiter} \citep[SO;][]{Muller2020} mission. Concurrently, the Slow Solar Wind Connection Solar Orbiter Observing Plan (Slow Wind SOOP; Yardley et al. 2023, submitted) was in operation during its second remote sensing window (2022 March 17 -- 22). This campaign used remote sensing and in situ instruments onboard SO to investigate the origin of the slow solar wind at open-closed magnetic field boundaries. The northwestern edge of the south polar CH was chosen as the target during 2022 March 17--18, which perfectly coincided with the time of the CME eruption and the merging event.

The merging of coronal holes and coronal dimmings has only been reported a few times \citep{Nitta2017, Gutierrez2018, Lorincik2021, Yardley2021a, Nitta2021}, and has never been studied spectroscopically due to their rare occurrence and transient nature. Fortunately, this merging event was well captured by various spacecraft, allowing us to analyse the merging in detail using high-resolution EUV and X-ray observations alongside the coronal spectroscopic measurements. In this paper, we investigate the changes in plasma dynamics inside the CH and the coronal dimming due to the merging and propose a possible process behind the merging of two open-field structures. The instrumentations and datasets used in this study and the technique used to identify the boundaries of the CH and coronal dimming are described in Section \ref{sec:obs}. The analysis results, including the evolution of CH and coronal dimming, plasma dynamics across four different timesteps during the merging event, and an example of a jet found inside the merged region are presented in Section \ref{sec:res}. Finally, we discuss the results and their implications in Section \ref{sec:disc} and summarise the manuscript in Section \ref{sec:sum}.

\section{Observations} 
\label{sec:obs}
On 2022 March 18, a filament channel and the southern polar coronal hole were observed on the solar disc. The filament partially erupted at $\sim$ 02:00 UT, resulting in a coronal mass ejection and a pair of coronal dimmings. The southeastern coronal dimming then expanded towards the southern polar CH due to the expansion of CME. Around 08:15 UT, the coronal dimming reached the northwestern tip of the CH boundary and appeared to merge with the CH, as observed using EUV and X-ray passbands.

\subsection{Coordinated Remote Sensing Observations}
The merging and evolution of CH and coronal dimming were observed using EUV images obtained from the Atmospheric Imaging Assembly \citep[AIA;][]{Lemen2012} onboard \textit{Solar Dynamics Observatory} \citep[SDO;][]{Pesnell2012}, and the Extreme Ultraviolet Imager \citep[EUI;][]{Rochus2020} onboard SO. SDO/AIA continuously observes the full solar disk in seven EUV passbands, with a plate scale of 0.6\arcsec\ per pixel and 12 s cadence. EUI consists of two telescopes, the Full Sun Imager (FSI) and High-Resolution Imager (HRI). FSI observes the full solar disc with a resolution of 4.4\arcsec\ per pixel and a cadence of 10 minutes. HRI offers a smaller field-of-view (FOV) of 1000\arcsec$\times$1000\arcsec\, but it has a high spatial and temporal resolution with a plate scale of 0.492\arcsec\ per pixel and a cadence of 5 s. HRI was operating from 10:15 UT to 11:15 UT on 2022 March 18, capturing the merged region. Observations from AIA were downloaded through the Virtual Solar Observatory (VSO) databases as level 1 data and processed using the \textit{aiapy} \citep{Barnes2020} Python library. For the EUI observations, we used calibrated level 2 EUI data from EUI data release 6 \citep{euidatarelease6}.

The line-of-sight (LOS) photospheric magnetograms, obtained from the Helioseismic and Magnetic Imager \citep[HMI;][]{Scherrer2012} onboard SDO, were used to study the magnetic field structure inside CH and coronal dimming. The full disc magnetograms were obtained from the HMI front 4096 $\times$ 4096 pixel CCD camera, which measures the Zeeman splitting of the Fe I 6173.3 \AA\ absorption line. The temporal resolution is 45 s with a plate scale resolution of 0.505\arcsec\ per pixel. The photon noise level is approximately 7 G \citep{Couvidat2016}. The HMI magnetograms were also downloaded from the VSO database and were coaligned to match AIA observations.

In addition to the EUV images, we used coordinated observations from the Extreme-Ultraviolet Imaging Spectrometer \citep[EIS;][]{Culhane2007} and the X-Ray Telescope \citep[XRT;][]{Golub2007} onboard \textit{Hinode} \citep{Kosugi2007}. Both instruments were observing the boundary of the south polar CH on 2022 March 17--18, as planned in the Hinode Operation Plan 434. On the day of the merging event, four EIS raster scans were obtained using two different EIS studies. The first study was HPW021VEL240x512v2\_b, a raster scan including a large variety of spectral lines with a FOV of 240\arcsec$\times$ 512\arcsec, in 2\arcsec\ slit steps lasting 3 hr 34 min. This study was run at 01:22 UT and 11:52 UT. Another study used was Atlas\_60, a full CCD spectrum scan with a FOV of 120\arcsec$\times$160\arcsec. This study uses a 2\arcsec\ slit with a total run time of 65 min. This study was run at 07:13 UT and 09:50 UT. The Doppler and nonthermal velocity can be derived from spectral data, with the velocity uncertainty of $\sim$ 5 km s$^{-1}$ for Doppler velocity and $\sim$ 20 km s$^{-1}$ for nonthermal velocity.

XRT was observing the southern polar CH boundary from 2022 March 17, 06:00 UT to March 18, 19:40 UT and continuously provided soft X-ray observations using the thin Al-poly filter, corresponding to a temperature of $\sim10^7$ K. The observations have a plate scale of 1\arcsec per pixel, and a FOV of 384\arcsec$\times$384\arcsec, except for the observation during 09:48--11:29 UT on March 18, where the resolution was resampled to 4\arcsec\ per pixel, with a FOV of 2106\arcsec$\times$2106\arcsec. 

EIS data files were downloaded from the UCL/MSSL database and were prepared using standard EIS routines in the SolarSoftWare Library \citep[SSW/IDL][]{Freeland1998}. Spectra were corrected for instrumental effects, including slit tilt, orbital variation, dark current, and warm/hot/dusty pixels. The XRT data were obtained from Hinode/DARTS managed by ISAS/JAXA and were prepared using the standard routine XRT\_prep.pro, also available in SSW/IDL.

\begin{figure*}[ht!]
    \begin{interactive}{animation}{Fig1Movie_edit.mp4}
    \includegraphics[width = 0.95\textwidth]{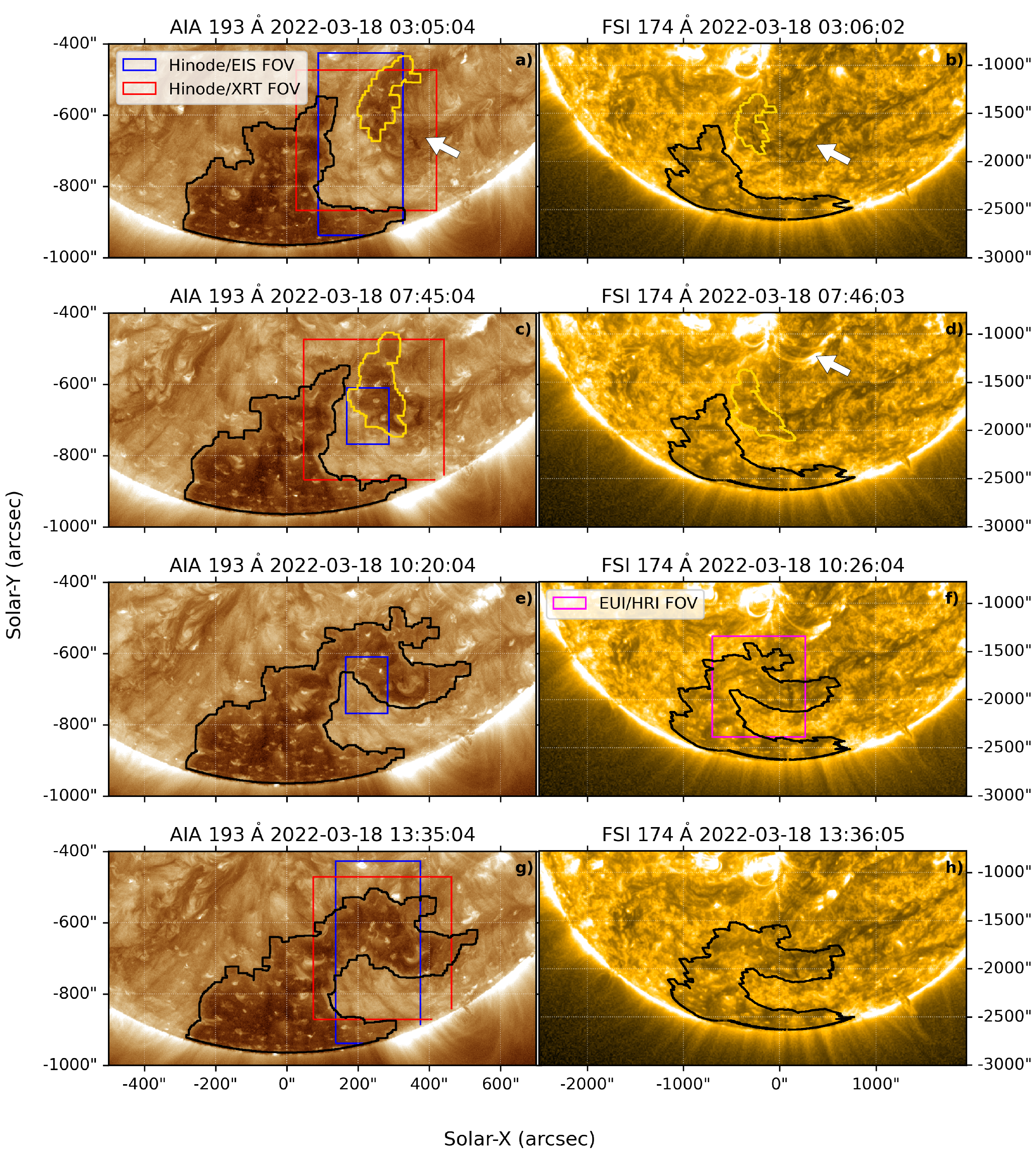}
    \end{interactive}
    \caption{SDO/AIA 193 {\AA} (left column) and EUI/FSI 174 {\AA} (right column) observations on 2022 March 18, capturing four stages of the merging between a southern polar coronal hole (CH) and a coronal dimming. The CH and coronal dimming boundaries are plotted in black and yellow, respectively. The FOVs of the coordinated observations are shown in coloured rectangles. The white arrows in panels a and b indicate the location of the filament channel, while the flare loops are indicated by the white arrow in panel d. The intensity maps have been enhanced using the Multiscale Gaussian Normalization method \citep{Morgan2014}. An animated version of this figure is available as Fig1Movie\_edit.mp4. The movie has a duration of 16~s and shows the evolution of the polar CH and the coronal dimming as observed from SDO/AIA and EUI/FSI from 00:00 UT to 14:00 UT.}
    \label{fig:AIA_EUI_CHevolution}
\end{figure*}

\subsection{Identifying Coronal Hole and Coronal Dimming Boundaries}
Coronal holes and coronal dimmings are best observed at a coronal temperature of $\sim10^6$ K. Therefore, we used the 193 {\AA} passband of SDO/AIA, and the 174 {\AA} passband of EUI/FSI and EUI/HRI, to track the global structure of the CH and coronal dimming. Note that SO was at an angle of 10$^{\circ}$ to the Sun-Earth line, allowing us to observe this event from two different viewpoints. 

The CH and coronal dimming boundaries were defined using SDO/AIA 193 {\AA} observations. For the CH, we implemented an intensity thresholding method, with the threshold intensity chosen to be 35\% of the solar disc median intensity \citep{Reiss2016, Hofmeister2017, Heinemann2018b}. For the coronal dimming, on the other hand, we used base-ratio images to enhance the relative changes in brightness \citep{Dissauer2018b}. The SDO/AIA 193 \AA\ observation at 01:30:04 UT, $\sim$30 minutes before the eruption started, was chosen as the reference image. Subsequent images were divided by the reference image to produce base-ratio images at each timestep. A thresholding technique was then used to detect coronal dimmings, with a threshold value of 0.35 \citep{Dissauer2018b}. After obtaining the initial boundaries for the CH and coronal dimming, we applied a Gaussian filter to each image to smooth the sharp edges and fill any holes inside the boundaries. The resulting smooth-edge boundaries of CH and coronal dimming for each time step were then projected onto other remote sensing observations. 
\section{Results} \label{sec:res}
\subsection{Evolution of Coronal Hole and Coronal Dimming}
Figure \ref{fig:AIA_EUI_CHevolution} shows the evolution of the southern polar coronal hole and the coronal dimming observed by SDO and SO at four different timesteps, corresponding to the time at the middle of each EIS study\footnote {Note that the distances of the individual spacecraft from the Sun meant that phenomena were observed at different times by the different spacecraft. Hence, to avoid confusion, we will use the time at Earth throughout this discussion.}. The boundaries of CH and coronal dimming are illustrated as black and yellow contours, plotted over SDO/AIA 193 {\AA} and EUI/FSI 174 {\AA} observations. This colour scheme is used throughout this paper. Each instrument’s FOVs are displayed as coloured rectangles. The blue rectangle is the FOV of Hinode/EIS, the red rectangle shows the FOV of Hinode/XRT, and the magenta rectangle shows the FOV of EUI/HRI. An animated version of Figure \ref{fig:AIA_EUI_CHevolution} shows the evolution of the polar CH and the coronal dimming from 00:00 UT to 14:00 UT.

From the SDO's point of view, the northern tip of the southern polar CH was close to the central meridian, with the northernmost edge at $y$ = -500\arcsec\ . The filament channel, indicated by the white arrows in Figure \ref{fig:AIA_EUI_CHevolution}a-b, was located $\sim$100\arcsec\ west of the CH. It appeared to have a U-shaped structure spanning at least half the size of the solar disk. Using photospheric LOS magnetograms from HMI, we found that the CH had net negative polarity, the same polarity as the eastern side of the filament channel.
\begin{figure}[t]
    \centering
    \includegraphics[height = 0.19\textheight, width=\columnwidth]{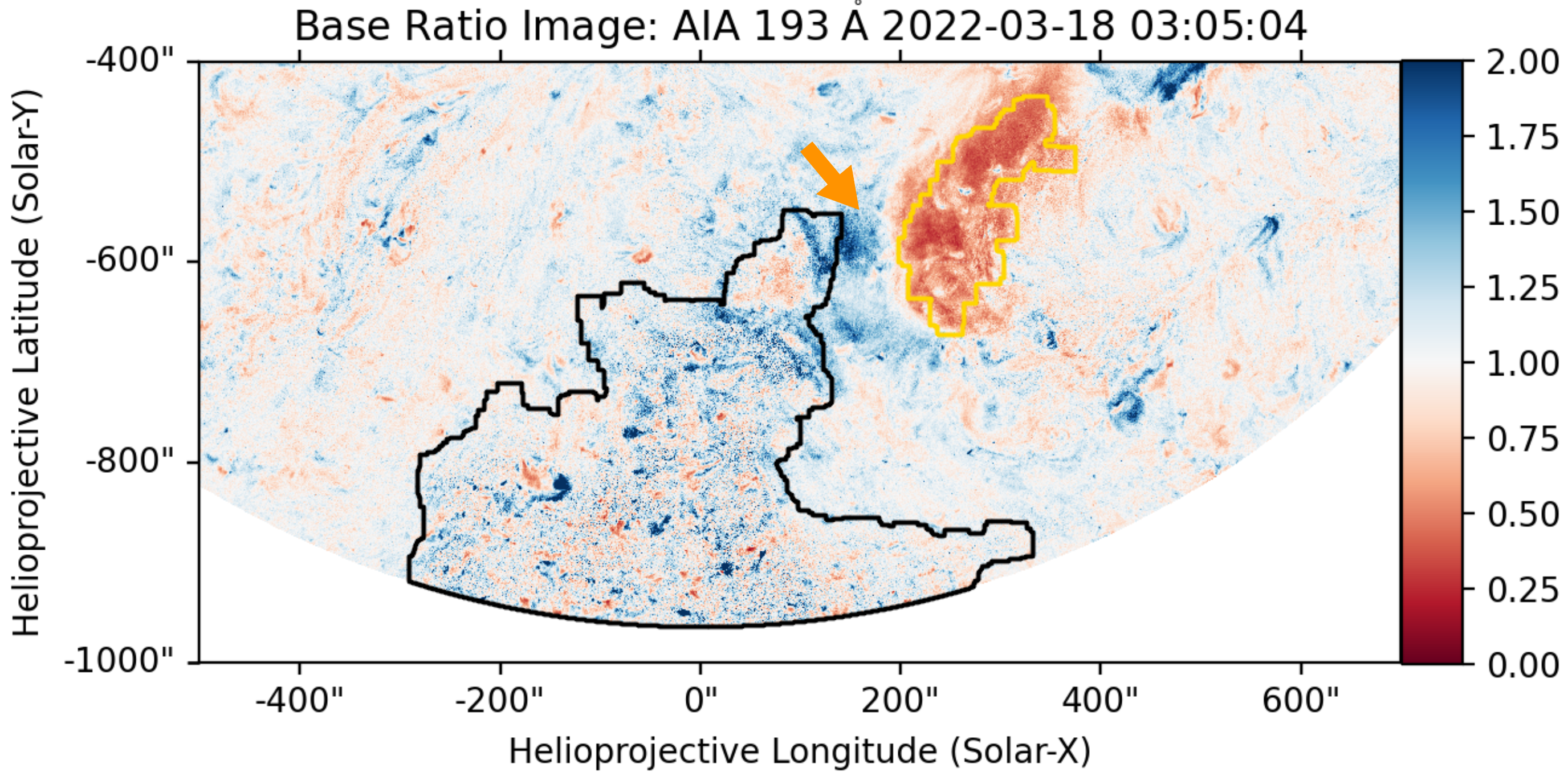}
    \caption{Base-ratio image produced from dividing SDO/AIA 193 {\AA} intensity at 03:05:04 UT by a reference image before the eruption (see text). Boundaries of the coronal dimming and the CH are plotted in yellow and black, respectively.  The yellow arrow indicates the region between CH and coronal dimming where the intensity enhancement (dark blue) is observed.}
    \label{fig:AIA_LBR}
\end{figure}
\begin{figure}[t!]
    \centering
    \includegraphics[width=\columnwidth]{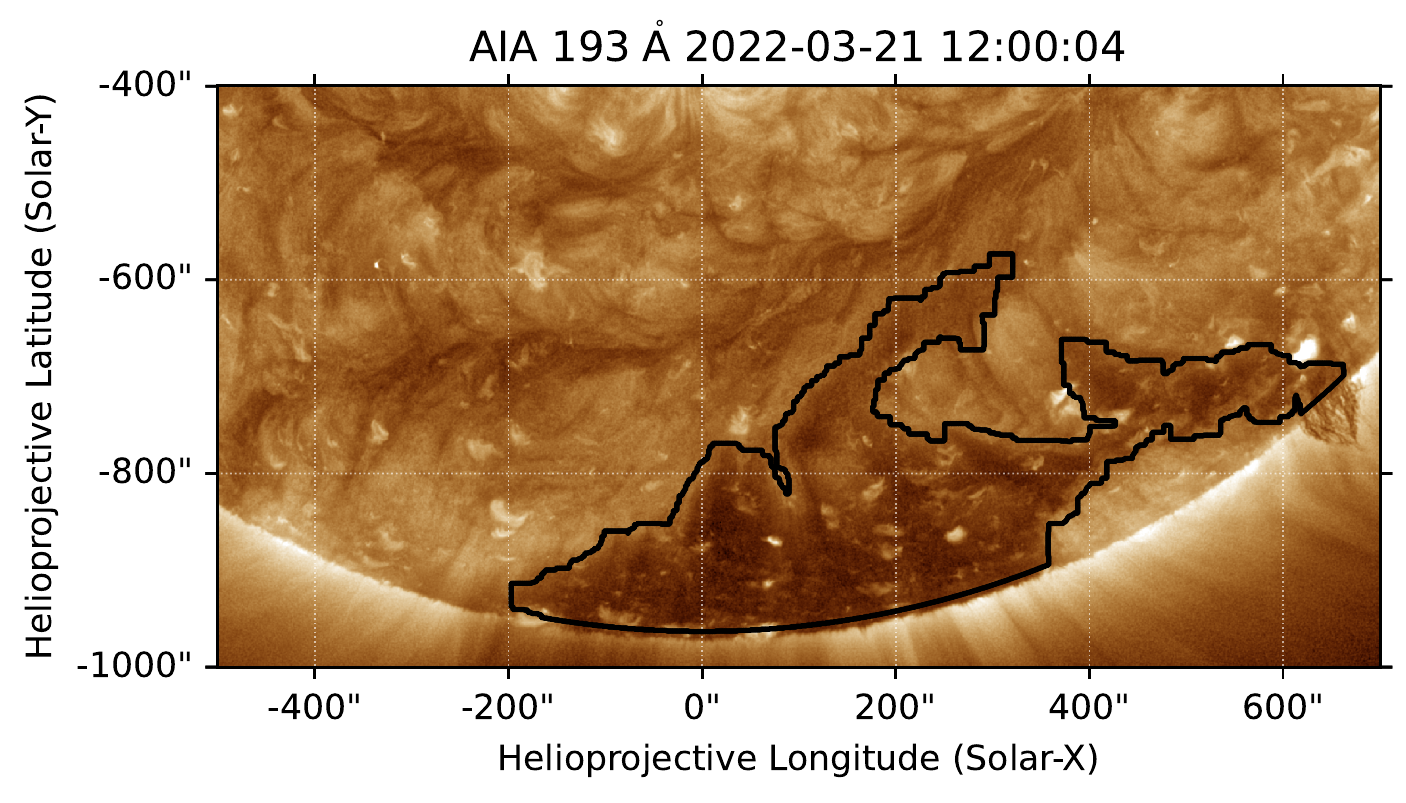}
    \caption{SDO/AIA 193 {\AA} observation on 2022 March 21, showing the boundary of the southern polar coronal hole three days after the merging occurred.}
    \label{fig:AIA_21March}
\end{figure}

 The filament partially erupted at the northern part of the U-shaped structure at $\sim$02:00 UT, resulting in a CME, flare loops, and coronal dimmings. The flare loops can be seen close to the central meridian from EUI/FSI 174 {\AA} at 03:36 UT, as indicated by the white arrow in Figure \ref{fig:AIA_EUI_CHevolution}d. The flare ribbons at the footpoints of the flare loops can also be observed using 304 \AA\ passband of EUI/FSI (not shown here). Shortly after the eruption had started, the coronal dimming started expanding in a southeastern direction towards the polar CH, accompanied by the expansion of the flare loop footpoints. Figure \ref{fig:AIA_LBR} shows a base ratio image which was used to define the boundaries of the coronal dimming. The coronal dimming can be observed as an area of decreased intensity, shown by red pixels. Figure \ref{fig:AIA_LBR} also shows the enhanced intensity (dark blue pixels) at the region between the southern polar CH and a coronal dimming, which are not evident in normal 193 {\AA} intensity observations. 
 \begin{figure*}[ht!]
    \centering
    \includegraphics[width=0.8\textwidth]{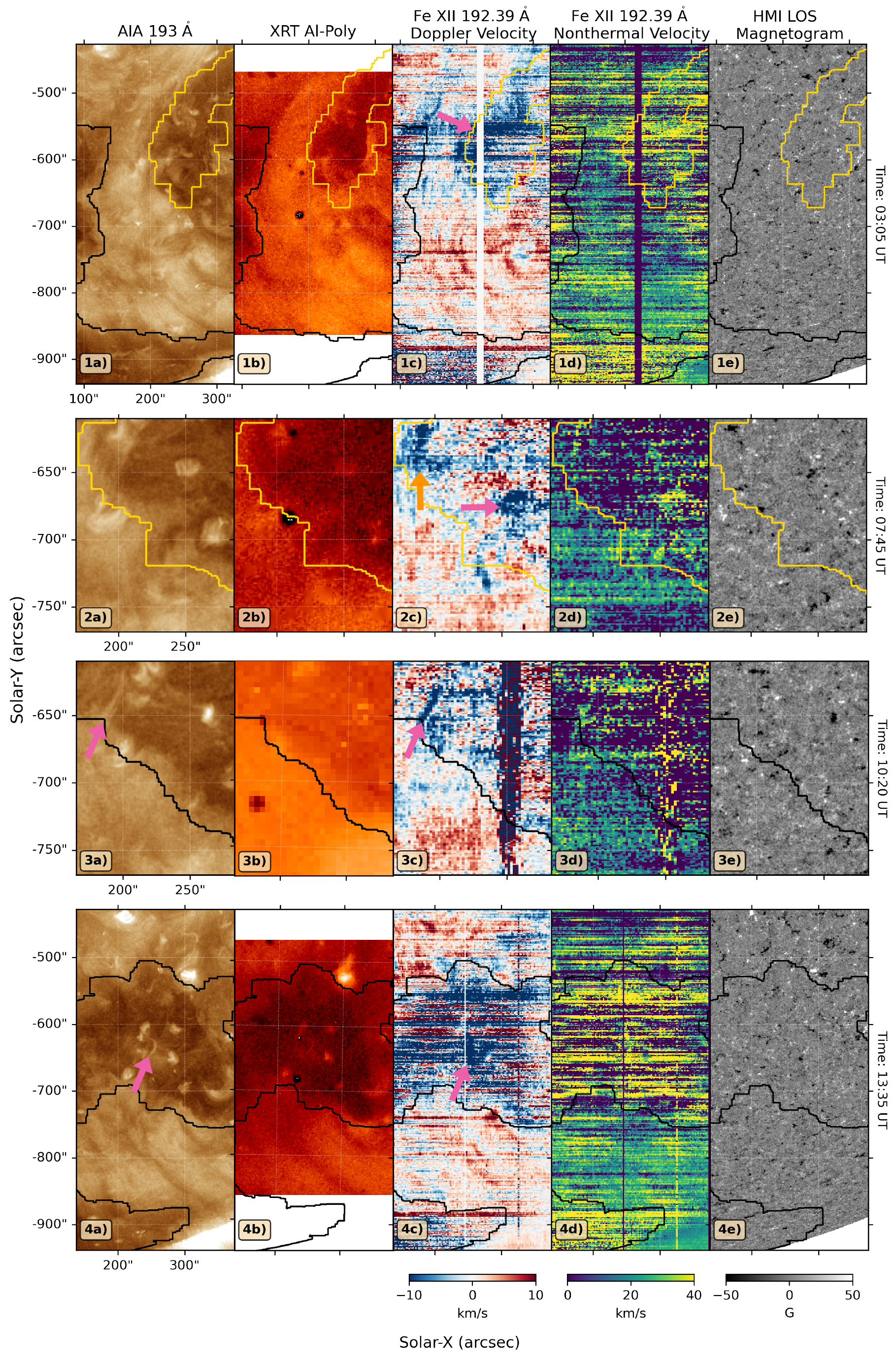}
    \caption{Plasma diagnostics of the coronal hole and coronal dimming at four EIS time steps. From left to right, the columns show a) SDO/AIA 193 {\AA} intensity, b) XRT Thin Al-Poly filter intensity, c) Fe XII 192.39 {\AA} Doppler velocity map, d) Fe XII 192.39 {\AA} nonthermal velocity map, and e) HMI line-of-sight photospheric magnetogram. The boundaries of CH and coronal dimming are shown in black and yellow, respectively. The colour scale of Doppler velocity maps was set to [-10,10] km s$^{-1}$. The nonthermal velocity maps are saturated at [0,40] km s$^{-1}$. HMI magnetograms are saturated at $\pm$ 50 G. The coloured arrows indicate the interesting upflow regions discussed in Section \ref{sec:res}.2.1.}
    \label{fig:Plasmadiag_4timestep}
\end{figure*}

 The coronal dimming reached the northwestern edge of CH at $\sim$07:30 UT before appearing to merge with the CH at $\sim$08:15 UT. Note that we define the merging as the moment the CH and coronal dimming boundaries appear as a single contour using the threshold method. Also, note that the stable part of the filament (lower arc) was included in the merged region on the western side beyond $x \sim$ 300\arcsec\ due to its low intensity comparable to the CH in SDO/AIA 193 {\AA} passband. However, the flare loop footpoints stopped their expansion at around 11:30 UT, 3 hrs after the merging occurred. According to the standard flare-CME model \citep[CSHKP model,][]{Carmichael1964,Sturrock1968,Hirayama1974,Kopp1976}, the expansion of flare loop footpoints indicates the ongoing reconnection process that simultaneously forms the flare loops and builds more magnetic flux into the expanding CME, which results in the expansion of coronal dimming. Therefore, we may infer that the coronal dimming was still expanding after it merged with the CH until at least 11:30 UT. 

 In summary, the EUV and X-ray observations on 2022 March 18 show the evolution of the coronal dimming and the southern polar CH as follows:
    \begin{itemize}
        \item 02:00 UT - The eruption started and the coronal dimming began its formation
        \item 03:36 UT - The flare loops can be seen from EUI/FSI 174 \AA\
        \item 08:15 UT - The coronal dimming merged with the southern polar CH
        \item 11:30 UT - The footpoints of flare loops stopped their expansion.
    \end{itemize}

Figure \ref{fig:AIA_21March} shows the merged region as seen from SDO/AIA 193 {\AA} on 2022 March 21, before the merged region moved away from the Earth-facing solar disc. Interestingly, this figure shows that this merged region persisted for at least three days after the merging occurred. The merged region then appeared to shrink in size and later disappeared on 2022 March 22, leaving behind only the southern part of the polar CH (the part of CH located approximately below $y \sim$ -800\arcsec\ ).

\subsection{Spectroscopic Data Analysis}
 Hinode/EIS was pointing at the western boundary of the CH, where the merging occurred. This allowed us to track and analyse the plasma dynamics at the time before, during, and after the merging of the CH and the coronal dimming by using four consecutive EIS raster scans. Although the merging of CHs and coronal dimmings have been reported in previous literature, this is the first time that such an event was observed spectroscopically for the entire process.

\subsubsection{Overview For Each EIS Raster}
Figure \ref{fig:Plasmadiag_4timestep} shows the Doppler velocity and nonthermal velocity maps obtained from fitting the Fe XII 192.39 {\AA} spectral lines for four EIS rasters, along with SDO/AIA 193 {\AA} intensity, XRT thin Al-poly filter intensity and HMI line-of-sight (LOS) magnetogram at the median time of each EIS scan (indicated on the right axis of each row). We will use the median time to represent the observation time of each study throughout the paper. The Fe XII 192.39 {\AA} spectral line was used because it is unblended, and the reference wavelength is close to the peak emission and temperature response of the SDO/AIA 193 {\AA} observation. All observations were aligned to the SDO/AIA 193 {\AA} intensity maps, with the CH and coronal dimming boundaries then overplotted. 

The first raster was taken at $\sim$03:05 UT, associated with the expansion of coronal dimming toward the east as the filament eruption was in progress. In this raster, we can see a large region of strong blueshift in the Doppler velocity map (see panel 1c in Figure \ref{fig:Plasmadiag_4timestep}), indicating a strong plasma upflow inside the coronal dimming region. 
This structured strong plasma upflow inside the coronal dimming has been reported in previous literature and usually has been interpreted as a result of rapid plasma evacuation onto expanding CME structure \citep[see, e.g.,][]{Harra2007, Attrill2010a, Tian2012, Veronig2019}. 

During the second raster at $\sim$07:45 UT, EIS observed the southeastern edge of the coronal dimming. During that time, the coronal dimming was located close to the southern polar CH boundary. Several bright points, previously located in the quiet Sun region, were found inside the boundary of coronal dimming (see panel 2a, 2b in Figure \ref{fig:Plasmadiag_4timestep}). There are two significant upflows regions in the Doppler velocity map, indicated by pink and orange arrows in panel 2c of Figure \ref{fig:Plasmadiag_4timestep}. The first region (pink arrow) is located between two bright points at (260\arcsec, -680\arcsec). Interestingly, this region had no observable signature in SDO/AIA 193 \AA\ and XRT observation, despite clear blueshift and enhanced nonthermal velocity. The second region (orange arrow) is located at (170\arcsec, -630\arcsec), close to a bright point and the interface between the CH and coronal dimming boundaries. Similarly, the source for this upflow is also uncertain. 

The third EIS raster observed the newly merged at $\sim$10:20 UT. A clear signature of a thin collimated jet can be seen at (190\arcsec, -650\arcsec), indicated by two pink arrows in panels 3a and 3c of Figure \ref{fig:Plasmadiag_4timestep}. This feature can be observed as a thin corridor of blueshifted plasma in the EIS Doppler velocity map and a collimated thread-like funnel in SDO/AIA 193 \AA\ observation. The base of upflow is located next to a bright point, which is the same bright point located next to the observed upflow indicated by the orange arrow in panel 2c of Figure \ref{fig:Plasmadiag_4timestep}.
\begin{figure*}[t!]
    \centering
    \includegraphics[width= 0.85\textwidth]{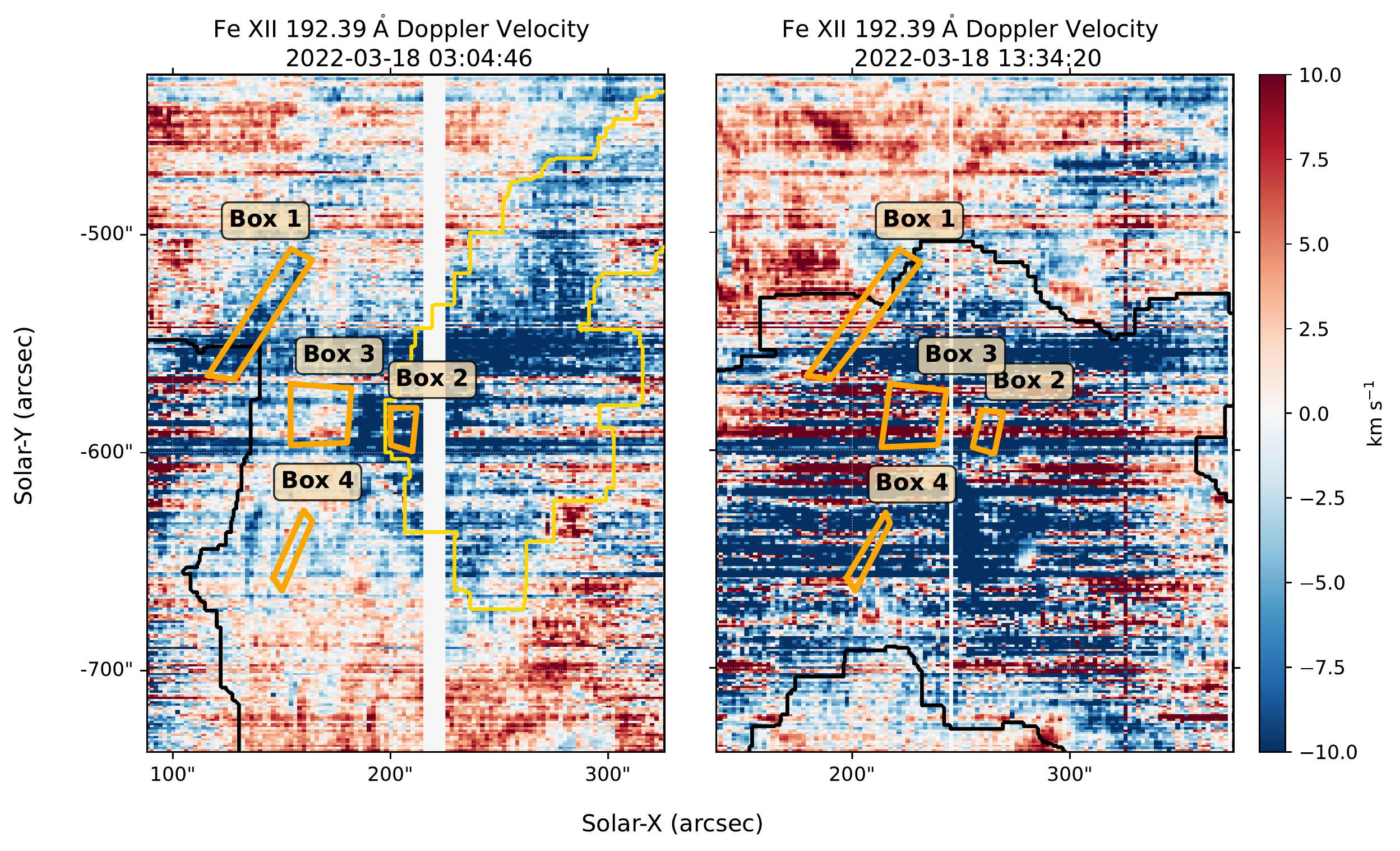}
    \caption{EIS Fe XII 192.39 \AA\ Doppler velocity maps obtained from EIS studies centred at $\sim$03:05 UT and $\sim$13:35 UT. The orange boxes indicate the approximate locations of regions where we averaged the spectral data to obtain the plasma dynamics shown in Table \ref{tab:EISresult_boxes}.}
    \label{fig:EISmap_boxes}
\end{figure*}

The fourth raster was run at $\sim$13:35 UT, 5 hours after the merging occurred. The Doppler and nonthermal velocity maps show that plasma flow became more disordered at this time, as indicated by the enhanced nonthermal velocity and the mixture of plasma upflow and downflow inside the merged region. However, this may be partially due to instrumental effects such as warm pixels. 
A relatively structured upflow region was observed at (250\arcsec, -650\arcsec), indicated by a pink arrow in panels 4a and 4c in Figure \ref{fig:Plasmadiag_4timestep}. It seems to be related to a breakout jet resulting from a mini-filament eruption at the bright point that occurred at $\sim$13:30 UT.

\subsubsection{Spatially Averaged Pixel Analysis}
We use the spatially averaging method described in EIS Software Note No.17 \footnote{Access through \url{https://vsolar.mssl.ucl.ac.uk/JSPWiki/Wiki.jsp?page=EISAnalysisGuide}} to obtain important plasma dynamic parameters across the four EIS rasters in several regions of interest. In doing this, we sacrifice the spatial resolution to enhance the signal-to-noise ratio of the low-emission coronal hole and coronal dimming.

First, we select four regions of interest, denoted as Box 1--4 in Figure \ref{fig:EISmap_boxes}. Box 1 is the thin upflow corridor at the north tip of the CH that transcended the boundary into the quiet Sun region. Box 2 is the strong upflow region inside the coronal dimming, 
Box 3 is the quiet Sun region between the CH and the filament channel before the merging event. Lastly, Box 4 is the jet's location observed in the EIS raster at 10:20 UT. The approximate locations of these boxes are shown in Figure \ref{fig:EISmap_boxes}. Next, we treat each box as a single pixel by averaging spectral data inside the box. We then derive the plasma parameters by fitting a Gaussian function to the averaged spectral data. Note that in some cases, we choose smaller regions at the same location with the boxes shown in Figure \ref{fig:EISmap_boxes} to obtain a better spectral fit.
\begin{table*}[ht!]
\resizebox{0.95\textwidth}{!}{%
\begin{tabular}{@{}ccccc@{}}
\toprule
Box Number (Feature) & EIS raster &
  Doppler Velocity (km s$^{-1}$) &
  Nonthermal Velocity (km s$^{-1}$) &
  \begin{tabular}{@{}c@{}}Secondary\\Component\end{tabular} \\ \midrule
\multirow{2}{*}{Box 1 (CH Edge)} & 03:05 UT & -18 & 28 & No  \\
                       & 13:35 UT & -22 & 31 & No  \\ \midrule
\multirow{2}{*}{Box 2 (Coronal Dimming)} & 03:05 UT & -26 & 20 & Yes \\
                       & 13:35 UT & -16 & 33 & No  \\ \midrule
\multirow{2}{*}{Box 3 (Quiet Sun)} & 03:05 UT & 0   & 25 & No  \\
                       & 13:35 UT & -3 & 31 & No  \\ \midrule
\multirow{4}{*}{Box 4 (Jet)} & 03:35 UT & -9  & 27 & No  \\
                       & 07:45 UT & -6  & 28 & No  \\
                       & 10:20 UT & -30 & 24 & Yes \\
                       & 13:35 UT & -23 & 34 & No \\ \bottomrule
\end{tabular}%
}
\caption{Results from EIS analysis using the averaged spectral data inside four boxes of interest. The table shows the Doppler and nonthermal velocities for each box at different times and whether the spectra have significant signatures of the secondary components in the line profile. Note that the merging occurred at $\sim$08:15 UT.}
\label{tab:EISresult_boxes}
\end{table*}
\begin{figure*}[ht!]
    \includegraphics[width=\textwidth]{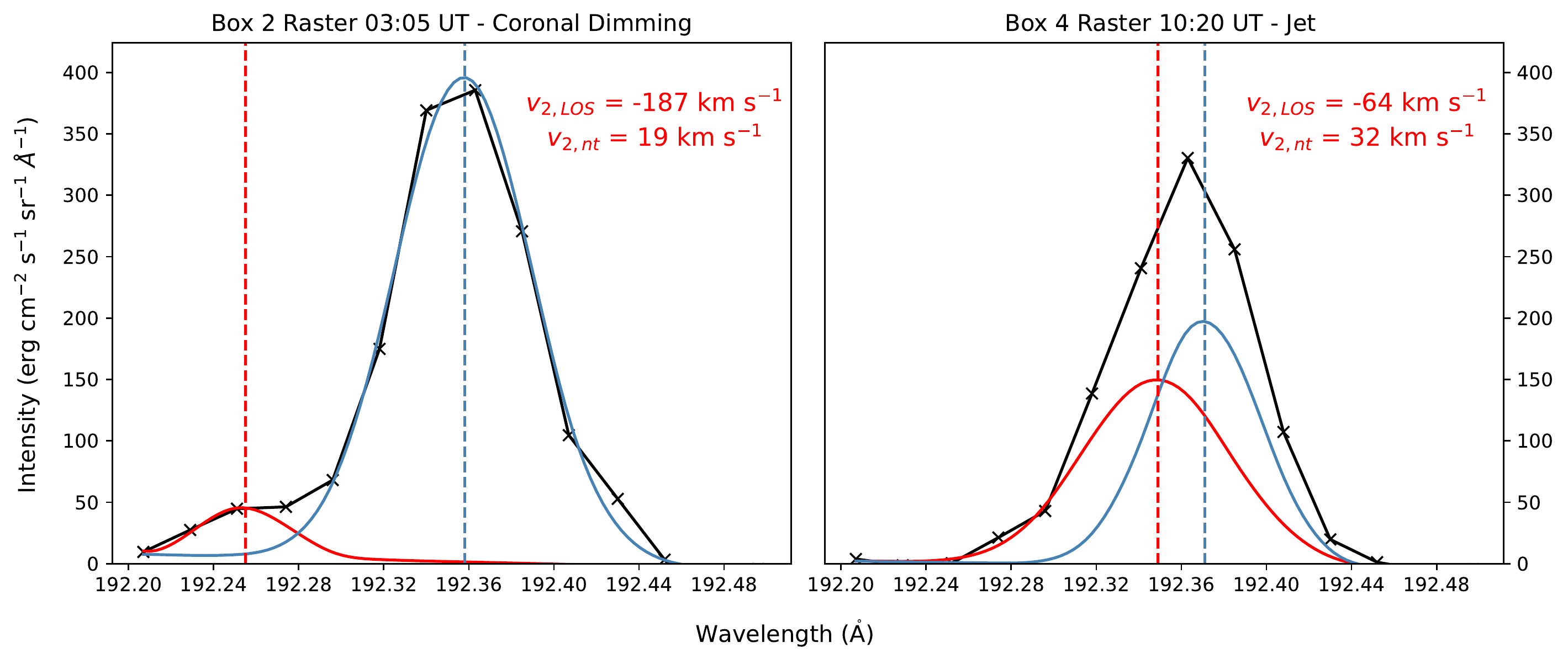}
    \caption{Fe XII 192.39 {\AA} emission line profiles obtained from the spatially averaged analysis for coronal dimming upflow (Box 2 in EIS raster at 03:05 UT; left panel) and jet upflow (Box 3 in EIS raster at 10:20 UT; right panel). Both line profiles show the signatures of blue-wing enhancement as indicated by the double gaussian fits, blue lines for the primary component and red lines for the secondary component. The calculated LOS velocity and nonthermal velocity of the secondary component are displayed in each panel.}
    \label{fig:EISprofile_BW}
\end{figure*}

 Table \ref{tab:EISresult_boxes} shows the results obtained from this spatially averaged pixel analysis using data from the Fe XII 192.39 {\AA} spectral line on all available four EIS rasters. Boxes 1, 2 and 3 are located outside the FOVs of the EIS rasters at 07:45 UT and 10:20 UT. Therefore, only Box 4 is observed on every raster. 
 
 In Box 1 (CH edge), we observed the upflow velocity of $\sim$20 km s$^{-1}$ and the nonthermal velocity of $\sim$30 km s$^{-1}$. These values are similar to the typical value of plasma dynamics inside CH observed using coronal spectral line \citep[e.g.,][]{Harra2015, Fazakerley2016}.    
 
 Boxes 3 and 4 are associated with the quiet Sun region before the merging, which later became the merged region of CH and coronal dimming and showed significant changes in plasma dynamics. Box 3 shows an increase in LOS velocity, as evident from the stronger redshift and blueshift pixels inside the box. The upflow profile changes from no upflow (on average) to upflow of $\sim$3 km s$^{-1}$. It also shows an increase in nonthermal velocity. The increase may be due to a change in the magnetic environment inside the region; the initially closed field region became more open because of the merging between CH and coronal dimming. The change in LOS velocity inside Box 4 is similar to Box 3, showing a more significant upflow when comparing between EIS raster at 03:05 UT and 13:35 UT, with a similar increase in nonthermal velocity. Furthermore, plasma dynamics remained almost the same in the EIS raster at 03:05 UT and 07:45 UT before changing drastically at the EIS raster at 10:20 UT, when there was a thin, collimated jet outflow inside the box.  
 
 Lastly, only Box 2 shows a decrease in LOS upflow velocity after the merging. The initial high upflow velocity in coronal dimming may be due to the plasma evacuation as the CME expanded. After the merging, the upflow velocity decreased to a level seen in the CH, while the nonthermal velocity increased similarly to other boxes.
 
By examining each line profile in detail, we see that some profiles have noticeable enhancements in the blue-wing region. This enhancement suggests that there are at least two components of unresolved plasma upflow along the LOS \citep{Tian2021}; namely, the background plasma flow (primary component) and high-speed upflow (secondary component). The strong secondary components have previously been observed in the upflows from active region boundaries \citep[e.g.,][]{Baker2021, Yardley2021b} and coronal dimming \citep{Tian2012, Veronig2019}. They can be identified by fitting a double gaussian instead of a single gaussian function to the profile. In our analysis, the averaged pixels from Box 2 in EIS raster at 03:05 UT and Box 4 in EIS raster at 10:20 UT show the significant secondary component in line profiles, i.e., the secondary component intensity is more than 10\% of the primary component \citep{Tian2021}. These line profiles correspond to the coronal dimming (Box 2) and the jet inside the merged region (Box 4). Both line profiles are shown in Figure \ref{fig:EISprofile_BW}; the coronal dimming in the left panel and the jet in the right panel. We also see comparable blue-wing enhancement in both regions in Fe XII 195.12 {\AA} and Fe XIII 202.04 {\AA} line profiles. 
\begin{figure*}[ht!]
    \begin{interactive}{animation}{Fig7Movie_edit.mp4}
        \includegraphics[width=0.95\textwidth]{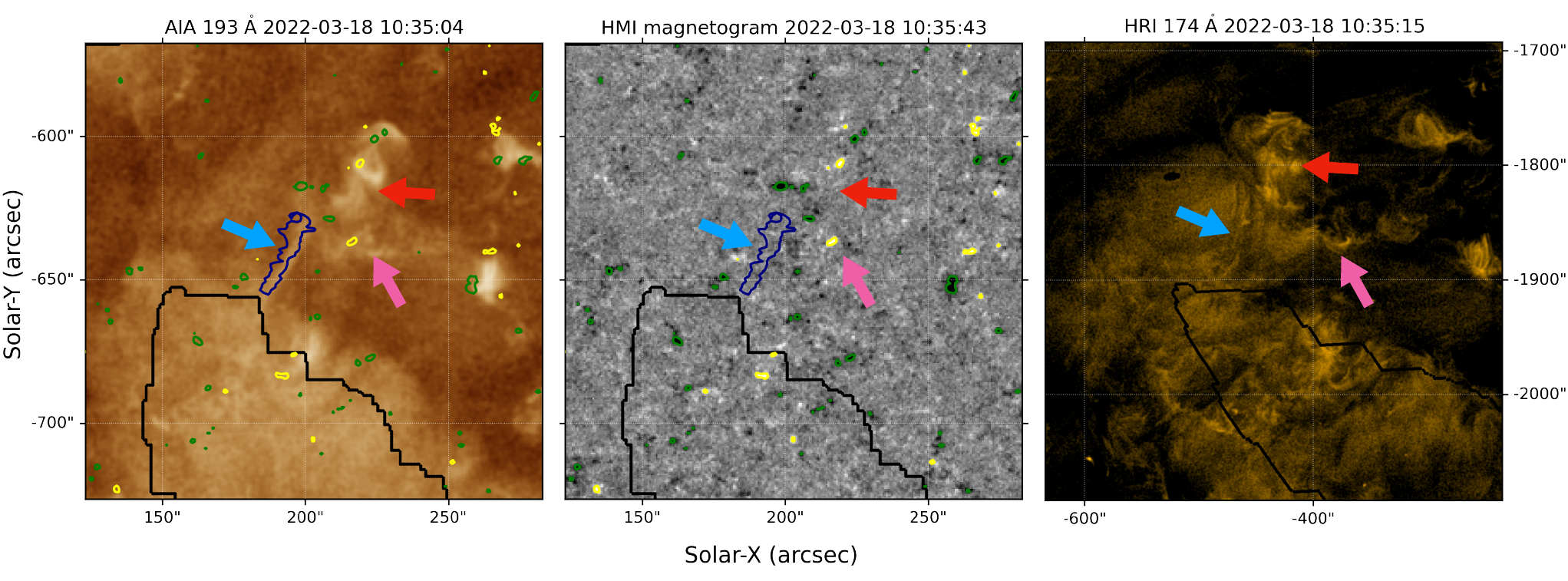}
    \end{interactive}
    \caption{A coronal jet inside the newly merged region as observed from SDO/AIA 193 {\AA} passband (left), HMI magnetogram (centre), and EUI/HRI 174 {\AA} passband (right) at 10:35 UT. The coloured arrows show the locations of the jet, the bright point and the adjacent small coronal loops. The contour of jet upflow observed in the EIS study centred at 10:20 UT is overplotted in dark blue. The green and yellow contours correspond to the magnetic field strengths of -50 and 50 G, respectively. An animated version of this figure is available as Fig7Movie\_edit.mp4. The movie has a duration of 24~s and shows the evolution of the jet as observed from SDO/AIA, SDO/HMI and EUI/HRI from 10:15 UT to 11:15 UT.}
    \label{fig:JetObs}
\end{figure*}

For the case of coronal dimming, the line profile looks like a single gaussian function with a bump on its blue-wing tail, indicating that the secondary upflow was weak but had a very high velocity. The secondary component has a LOS velocity of -187 km s$^{-1}$, higher than the reported value of CME upflows in the range of [50--150] km s$^{-1}$ \citep{Tian2012, Veronig2019}. We interpret that the high-speed plasma upflow is a result of impulsive plasma outflow along the expanding magnetic field structure \citep{Harrison2000,Harra2007,Veronig2019} rather than the plasma refilling the coronal dimming after the eruption \citep{Tian2012} due to the extended coronal dimming lifetime. The nonthermal velocity is 19 km s$^{-1}$, lower than the primary upflow. 
For the jet inside the merged region, the secondary component has a LOS velocity of -64 km s$^{-1}$. It also has a nonthermal velocity of 32 km s$^{-1}$, higher than the primary upflow. The secondary components for both features can be observed in only one EIS raster, indicating the transient nature of high-speed plasma outflows.

\subsection{A Coronal Jet at 10:30 UT}
X-ray and EUV observations show several coronal jets and brightenings of bright points during this merging event, especially in the interface region between the coronal hole and coronal dimming (see panels 2a, 2b, 3a and 3b of Figure \ref{fig:Plasmadiag_4timestep}). These bright points initially resided in the quiet Sun region between the CH and the filament channel, which later turned into the merged region.

An example of a coronal jet is shown in Figure \ref{fig:JetObs}. We selected this jet as an example because it was well observed in SDO/AIA, EUI/HRI and Hinode/EIS Doppler velocity map. Figure \ref{fig:JetObs} displays a thin-collimated beam of plasma observed by SDO and SO at $\sim$10:35 UT. The jet is located around (180\arcsec, -650\arcsec) in helioprojective coordinates as seen from SDO (blue arrows), close to the boundary of the post-merging CH and the bright point (red arrows). It also corresponds to the thin upflows corridor observed in the EIS raster at 10:20 UT (indicated by a pink arrow in panel 3c of Figure \ref{fig:Plasmadiag_4timestep}), which is shown as the blue contours in Figure  \ref{fig:JetObs}. 
The jet became visible in SDO/AIA 193 \AA\ at $\sim$10:00 UT. Note that the jet formation might start earlier but is not noticeable due to the brighter background emission. The jet initially had a weak emission and consisted of several thin upflows threads, with the jet’s base adjacent to the eastern edge of the bright point. It then appeared brighter and became more collimated into a thin beam around 10:25 UT. Besides the coronal jet, we also observed brightenings of coronal loops in the bright points and the adjacent loops under it (pink arrows in Figure \ref{fig:JetObs}) at $\sim$10:30 UT. Finally, the jet slowly dispersed and completely disappeared at $\sim$11:00 UT, marking a visible lifetime of $\sim$1 hr. 

The jet can be classified as a standard jet due to its similar configuration (long-thin spire) to the model proposed by \citet{Shibata1992} and the lack of emission in the 304 {\AA} passband \citep[cf.][]{Long2023}. Using the spectroscopic analysis discussed earlier, this jet had a LOS velocity of -64 km s$^{-1}$ with a nonthermal velocity of 32 km s$^{-1}$ (see right panel on Figure \ref{fig:EISprofile_BW}). 

This jet is not clearly visible against brighter background emission in EUI/HRI 174 {\AA} observations. However, we notice that the background emissions seem to have an outflowing motion from the bright point that the jet originated from, suggesting that they were coronal plumes, hazy ray-like transient structures commonly found in CH. Coronal plumes can be frequently observed using white light and SDO/AIA 171 {\AA} passbands and may also contribute to supplying mass and energy to solar wind \citep[see review by][and references therein]{Poletto2015}. The high resolution of EUI/HRI can resolve the motion of closed loops inside the bright point and surrounding area, which can be related to the reconnection process powering the jet. The observations also reveal numerous small-scale, short-lived brightenings that are similar to ‘campfires’ \citep{Berghmans2021} observed in the quiet Sun region.

The photospheric magnetogram (the middle panel in Figure \ref{fig:JetObs}) shows the minority polarity (positive, yellow contour) of the bright point surrounded by background majority polarity (negative, green contour) in the merged region. In our scenario, this configuration resembles dipoles embedded in a uniform open field, which is often associated with a 3D fan-spine structure of closed loops with a domed separatrix surface that separates closed and open magnetic field regions \citep[e.g.,][]{Kumar2019}. It is also frequently used as the initial setup for simulating coronal jet formation \citep{Yokoyama1995, Pariat2009, Wyper2018a}. In these setups, the interchange reconnections between the open background field and closed loops are responsible for allowing the reconnection exhaust to escape as jets along the newly open field lines, which is compatible to our observations.  

Note that there were multiple brightenings of bright points and jets inside the merged region other than this jet in Figure \ref{fig:JetObs}. These jets can be different from the example jet discussed above. For example, the jet observed at $\sim$13:30 UT (indicated by pink arrows in panel 4a and 4c in Figure \ref{fig:Plasmadiag_4timestep}) was likely to be a breakout jet corresponding to a mini-filament eruption from the bright point, as the EUV observations show the eruption of a small coronal loop with the brightenings observable in 304 \AA\ passband.

\section{Discussion} \label{sec:disc}
Our observations show that the coronal dimming intruded the boundary of the southern polar coronal hole, resulting in the merging of two open field structures. The merged region persisted for over 72 hr (see Figure \ref{fig:AIA_21March}), significantly longer than the typical coronal dimming lifetime of 3--12 hr \citep{Reinard2008}. The merged region then shrank and disappeared alongside the northern part of CH, suggesting that the coronal dimming became indistinguishable from the polar CH. If there were no interactions between the coronal dimming and the CH, the northern part of the CH should still exist following the disappearance of the coronal dimming. The spectroscopic observations from Hinode/EIS show that the upflow speed inside the coronal dimming region decreased to a roughly similar level with the CH after the merging, with no significant sign of a secondary Gaussian component, indicating an absence of superposed high-speed plasma upflow associated with the coronal dimming (see Table \ref{tab:EISresult_boxes}). The nonthermal velocities in every region included in the analysis were enhanced after the merging, further suggesting that there were underlying mechanisms contributing to the broadening of the line profile.

\subsection{Component Reconnection as Merging Mechanisms} 
\begin{figure}[ht!]
    \includegraphics[width = \columnwidth]{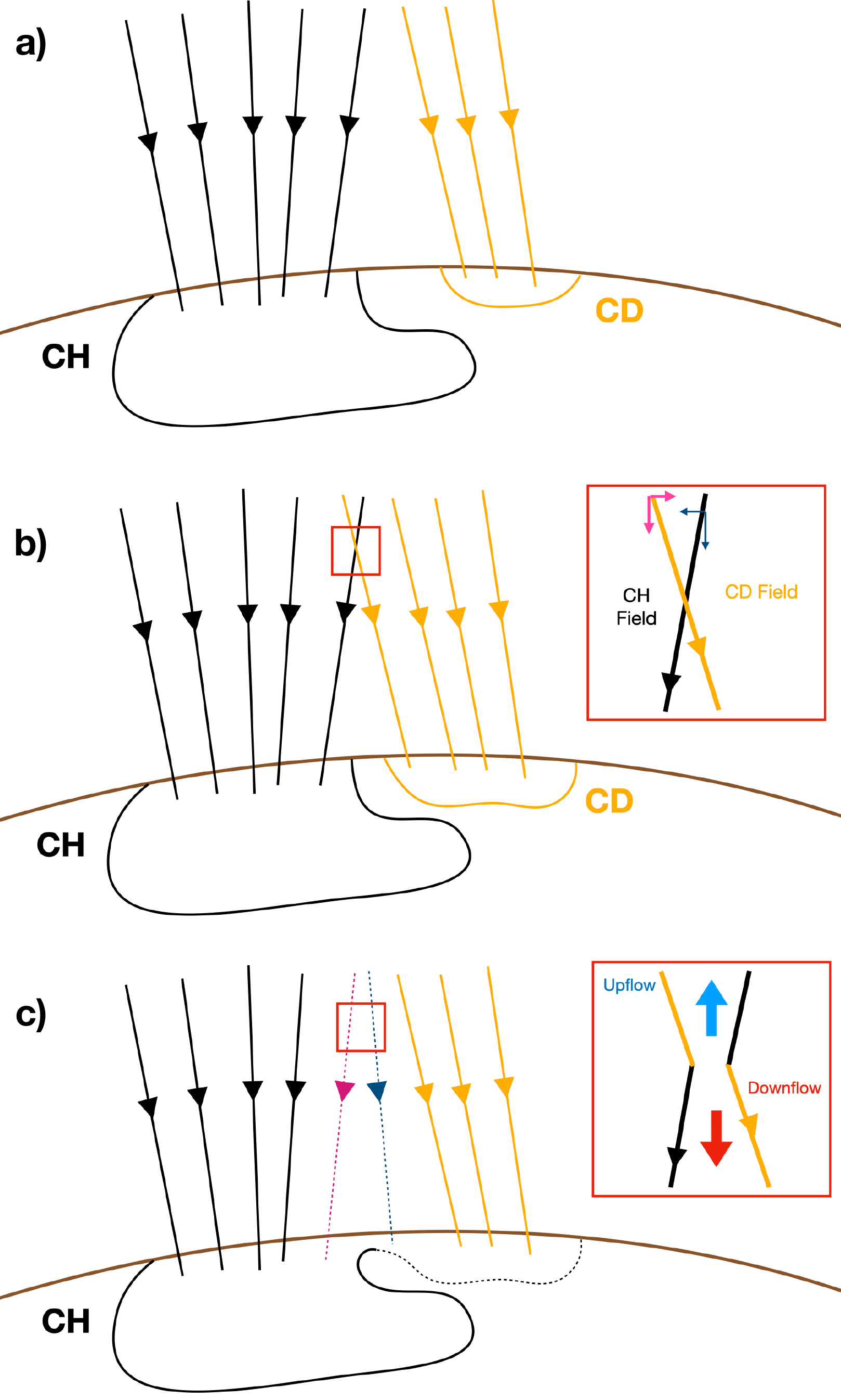}
    \caption{A 2D diagram illustrating the merging process of a coronal hole (CH, black) and a coronal dimming (CD, orange). (a) Initially, the CH and the coronal dimming were further apart. Coronal dimming field lines (orange lines) were at an angle to the CH field lines (black lines). (b) The coronal dimming then expanded towards the CH boundary. The field lines from both structures reconnected at the interface region. The zoomed-in view of the reconnection site is shown in the red square. The CH and coronal dimming magnetic field lines had an anti-parallel component, allowing component reconnection to occur. (c) The footpoints of CH and coronal dimming field lines were switched after the reconnection, as indicated by newly reconnected field lines (blue and pink). The mixture of plasma upflows and downflows observed after the merging may also be due to plasma outflows from the reconnection site.}
    \label{fig:Compreconnect}
\end{figure}

We propose that the main interaction involved in this merging process is the footpoint switching between open field lines in the CH and coronal dimming. Since both features had dominant negative polarity, we suggest that component reconnection was responsible for this process. Component reconnection is magnetic reconnection in which the magnetic field lines are not anti-parallel to each other, but there is still some angle between reconnecting field lines (i.e., one component of the field lines is anti-parallel). This type of reconnection can occur in unipolar magnetic field regions and can be found in various places and scales throughout the solar atmosphere \citep[e.g.,][]{Parker1983,vanDriel-Gesztelyi2012,Rappazzo2012,Tenerani2016,Antolin2021,Chen2021}. 

In this scenario, the main driver for the component reconnection between the CH and the coronal dimming is the extensive lateral expansion of the CME. Figure \ref{fig:AIA_LBR} shows the brightening at the region between coronal dimming and the southern polar CH in the base-ratio image. We speculate that this brightening may be a result of the plasma compression inside the overlying field arcade above the expanding CME since the overlying arcade may not be able to expand further east due to the open field of the coronal hole. Hence, the brightening implies that the lateral expansion of CME is powerful enough to compress the coronal plasma and potentially drive the reconnection process. 

Figure \ref{fig:Compreconnect} shows a diagram explaining the component reconnection during our merging event. Initially (panel a), the CME resulted in a coronal dimming which contained the newly open field. The coronal dimming field lines (orange lines) were at an angle with the field lines of nearby polar CH (black lines) due to the expansion of the CME. The boundaries of coronal dimming then reached the CH (panel b). Since at least one component of magnetic field lines from each region was anti-parallel, the component reconnection between coronal dimming and coronal hole magnetic fields could take place at the boundary interface, as shown in the diagram inside the red rectangle in panel b. 

Reconnection reconfigured the field lines and resulted in footpoint switching (panel c); the post-reconnection coronal dimming field (pink line) is rooted in the CH, and the CH field (blue line) is rooted in the coronal dimming. This made the boundaries of both regions overlap with each other, resulting in the merging and allowing the coronal dimming to intrude on the boundary of the CH. 

This scenario can explain why the coronal dimming was still able to expand after the merging occurred (as indicated from the expansion of flare loops footpoints in Figure \ref{fig:AIA_EUI_CHevolution}) since the ongoing footpoint switching allows the coronal dimming to keep expanding further inside the coronal hole. The footpoint switching might also distribute the CH fields throughout the merged region, which may help make the CH and coronal dimming indistinguishable and maintain the merged region's structure.  The plasma outflow due to the component reconnection is a plausible explanation for the mixture of disordered plasma upflow and downflow and the enhancement of nonthermal velocities observed after the merging in EIS Raster 13:35 UT (see panel 4c and 4d of Figure \ref{fig:Plasmadiag_4timestep}).

We hypothesise that the component reconnection started at a higher altitude and moved to a lower height as the merging progressed. The reconnection occurred continuously until the expansion of coronal dimming stopped, at which point there was no further movement of field lines that could drive the reconnection process.

\subsection{Jets and Upflows Inside Coronal Dimming and the Merged Region}

Using the broadband EUV and spectroscopic observations, we found numerous signatures of coronal plasma outflows throughout the merging processes, either as coronal jets in EUV observations or blueshift (upflow) regions in the Doppler velocity maps.

These events were more apparent after the merging, which could be due to the darker background. However, we believe that the main reason is the enhanced reconnection due to the change of the magnetic configuration caused by the expanding coronal dimming and the subsequent merging with CH. Before the eruption, these bright points resided in the quiet Sun region between the southern polar CH and the filament channel, where the magnetic configuration mainly consisted of closed magnetic loops. The erupting CME opened up the field lines, resulting in the observed coronal dimming, which later expanded to the region where bright points were located. This drastically changed the magnetic configuration from a closed environment to an open one, allowing more interchange reconnections and jet outflows.

In our observations, the open fields from the coronal dimming expanded across the bright point where the jet occurred. This scenario is analogous to the simulation setup for studying jets and bright point brightenings due to the moving magnetic elements in a CH by \citet{Wyper2018a}. In the first configuration in the study, they simulated a moving embedded minor polarity inside the background field of the net majority polarity, inducing the shear flow at the separatrix surface, which is similar to our scenario in the moving minority polarity frame. The results from the simulation indicated quasi-periodic weak bursts of interchange reconnection, which resulted in the brightening of closed loops and weak outflow. The outflow resembles a standard jet (i.e., long thin spire with no emission in the 304 {\AA} passband) with an outflow speed of around 60 km s$^{-1}$.

An example jet in Figure \ref{fig:JetObs} has a LOS velocity of -64 km s$^{-1}$ and has no signature in 304 {\AA} passband, which agrees with the results of the simulation by \citet{Wyper2018a}. Therefore, we believe that this jet corresponds with the simulation setup. However, the main differences between our events and the simulation are the driving force of the magnetic field movement and the angle of the background open field. The simulation drove the magnetic elements through the vertical field using photospheric flows, in contrast with the angled field passing through bright points due to the expansion of coronal dimming seen here. 

Although we reported the source of several upflows in the results, there are still some upflows whose sources were less certain, such as the two upflow regions inside the coronal dimming region observed in the EIS raster 07:45 UT (see the coloured arrows in panel 2c of Figure \ref{fig:Plasmadiag_4timestep}).  

The first upflow region (pink arrow) is located well inside the coronal dimming, showing a strong blue-shifted plasma with no EUV and X-Ray emission counterpart. One possible explanation is that these outflows are escaping plasma alongside expanding structure caused by the eruption, similar to CME-induced upflows observed in the EIS raster at 03:05 UT (panel 1c of Figure \ref{fig:Plasmadiag_4timestep}). However, due to its localised shape and position close to two bright points, the upflow could also be a structure resembling a `dark jet’ inside CHs discovered in a study by \citet{Young2015}. The second outflow region (orange arrow) is located at the interface region between the CH and coronal dimming and close to the bright point, which was the source of the jet in Figure \ref{fig:JetObs}. Hence, the upflows could also be a jet originating from the bright point. However, the brighter quiet Sun background in both EUV and X-ray observation made jet identification challenging. This upflow can also be caused by reconnection at the boundary region between CH, coronal dimming, and quiet Sun, either through interchange reconnection or component reconnection between open fields. 

\subsection{Relationship between Coronal Holes and Coronal Dimmings}
Coronal dimmings are often called transient coronal holes due to their similar appearance. However, they can be distinguished by their difference in lifetime and formation process. Coronal dimmings arise from solar eruptions and have a lifetime on a scale of hours, while CHs are thought to form more gradually and usually persist for several Carrington rotations. However, recent observations indicate that solar eruptions could be another way to form CHs \citep{Liu2007, Gutierrez2018, Heinemann2018b}, suggesting that some CHs may actually be coronal dimmings with unusually long lifetimes. Following this train of thought, we may infer that long-lived coronal dimmings and CHs are almost identical. Our observations show that the coronal dimming (merged region) persisted for more than 72 hr after its formation and became indistinguishable from the CH after the merging. This provides another example of a CH arising from coronal dimming and supports the idea that coronal dimmings and CHs are similar structures.

The recovery of coronal dimmings is thought to be facilitated by interchange reconnection with the surrounding quiet Sun regions \citep{Attrill2008}, which disperses the concentration of open field inside the coronal dimming and allows plasma to fill in the dimming region. Therefore, the speed of coronal dimming recovery may be related to the surrounding magnetic field environment. In a unipolar region, the coronal dimming recovery is expected to be slower due to fewer closed loops to reconnect with open fields. In our observation, the coronal dimming was located near the southern polar CH and later merged with each other. Since the coronal dimming and the CH had the same net magnetic polarity, this created a large, strong unipolar region which could slow the dispersion of the open fields to surrounding quiet Sun regions, resulting in an unusually long lifetime of coronal dimming that later became indistinguishable from CH. Several mergers of CHs and coronal dimmings and the conversions from coronal dimming to CH were reported by \citet{Gutierrez2018}, and all of them were found in unipolar magnetic environments. Therefore, we speculate that if eruptions occur near unipolar regions, the resulting coronal dimmings will likely have longer lifetimes and could later turn into isolated CHs or merge with a neighbouring CH.

Like CHs, coronal dimmings have been suggested as a potential source of the solar wind due to observed outflows and open-field configuration \citep{Thompson2000, Harra2007, Mcintosh2007}. \citet{Lorincik2021} reported evidence of solar wind from a coronal dimming as continuous outflow funnels were observed in 171 \AA\ and 193 \AA\ passbands of SDO/AIA. They lasted more than five hours and had a velocity of up to 70--140 km s$^{-1}$. The derived outflow speed agrees with the coronal dimming upflow speed obtained from spectroscopic observation in our work and previous literature \citep{Tian2012, Veronig2019}, and has a similar outflow pattern and velocity with neighbouring CH. Our result suggests that coronal jets and coronal plumes can originate from bright points inside the coronal dimming and could contribute to solar wind. The component reconnection between open fields, which could occur during our merging scenario, can also be an important process for solar wind acceleration \citep{Viall2020}. However, post-observation connectivity analysis using the magnetic connectivity tool \citep{Rouillard2020} reveals that it was unlikely that SO was magnetically connected with the western boundary of CH during the merging event. Therefore, the solar wind streams arising from this region were likely not observed by the spacecraft.

 \section{Conclusion} \label{sec:sum}

In this paper, we report and analyse the merging of the southern polar coronal hole with a coronal dimming resulting from a coronal mass ejection on 2022 March 18 during the Slow Wind SOOP coordinated observation campaign. The coordinated observation allowed us to analyse the merging process of two open-field structures in detail by combining high spatial and temporal resolution EUV and X-ray observations with the coronal spectroscopic data. The main results and discussion can be summarised as follows.
\begin{enumerate}
    \item The merging of the polar coronal hole and a coronal dimming resulted in the merged region that persisted for more than 72 hours, significantly longer than a typical lifetime of coronal dimmings. The cotemporal disappearance of the northern part of the coronal hole alongside the coronal dimming indicates that the coronal dimming and the coronal hole became indistinguishable from one another.
    \item The plasma dynamics were observed to become more disordered after the merging of the coronal hole and the coronal dimming, as evident in the mixing of upflow and downflow throughout the region and the overall enhancement of nonthermal velocities. Using a spatially averaged pixel analysis, we find that the upflow profile inside the coronal dimming and quiet Sun regions became more similar to the coronal hole upflow after the merging. In particular, the upflow velocity inside the coronal dimming changed from the order of hundred km s$^{-1}$ initially to the order of ten km s$^{-1}$ after the merging occurred.
    \item Component reconnection is considered the primary driving process in the merging of these two structures. The open fields from the coronal hole and the coronal dimming reconnect and exchange their footpoints, allowing the coronal dimming to intrude into the coronal hole region and making the coronal hole and the coronal dimming indistinguishable. 
    The mixture of plasma upflow and downflow observed after the merging may also be due to the reconnection.
    \item Several bright point brightenings and coronal jets are found inside the coronal dimming and the merged region throughout the observation period, suggesting ongoing reconnection processes. A study of the coronal jet inside the merged region shows that this jet may be driven by the interchange reconnection between the closed loops in a bright point and the expanding open fields in the coronal dimming.
\end{enumerate}
Coronal dimmings are traditionally called transient coronal holes due to their resemblance to one another. However, we provide evidence that they are also closely related, as coronal dimmings can merge with coronal holes and become the same structure. Their similarities also suggest that coronal dimming may play an important role in the formation of coronal holes. Further studies are needed to understand what happens during the merging of two open-field structures and the potential consequences in the heliosphere, such as precise magnetic field extrapolation during the merging, the study of plasma composition evolution throughout the merging process, and the in-situ observation of the outflows from the merged region. Our study also emphasises the importance of coordinated solar observations as a powerful means to reveal the mysteries of the Sun. By combining several observations from various spacecraft, we can improve our understanding of the coronal features, their magnetic field structures and their influence towards the heliosphere in the form of CME or solar wind.

\begin{acknowledgments}
We are grateful to Lidia van Driel-Gesztelyi, Gherado Valori and David Brooks for their helpful discussions which improves the content of this paper. Solar Orbiter is a space mission of international collaboration between ESA and NASA, operated by ESA. The EUI instrument was built by CSL, IAS, MPS, MSSL/UCL, PMOD/WRC, ROB, LCF/IO with funding from the Belgian Federal Science Policy Office (BELSPO/PRODEX PEA 4000134088); the Centre National d’Etudes Spatiales (CNES); the UK Space Agency (UKSA); the Bundesministerium f\"{u}r Wirtschaft und Energie (BMWi) through the Deutsches Zentrum f\"{u}r Luft- und Raumfahrt (DLR); and the Swiss Space Office (SSO). SDO data are courtesy of NASA/SDO and the AIA, EVE, and HMI science teams. Hinode is a Japanese mission developed and launched by ISAS/JAXA, collaborating with NAOJ as a domestic partner, NASA and STFC (UK) as international partners. Scientific operation of the Hinode mission is conducted by the Hinode science team organized at ISAS/JAXA. This team mainly consists of scientists from institutes in the partner countries. Support for the post-launch operation is provided by JAXA and NAOJ (Japan), STFC (UK), NASA (USA), ESA, and NSC (Norway). N.N. is supported by STFC PhD studentship grant ST/W507891/1 and UCL Studentship. D.M.L. is grateful to the Science Technology and Facilities Council for the award of an Ernest Rutherford Fellowship (ST/R003246/1). D.B. is funded under STFC consolidated grant number ST/S000240/1. S.L.Y. would like to thank NERC for funding via the SWIMMR Aviation Risk Modelling (SWARM) Project, grant number NE/V002899/1, and STFC via the consolidated grant number STFC ST/V000497/1. A.W.J. was supported by a European Space Agency (ESA) Research Fellowship and acknowledges funding from the STFC consolidated grant ST/W001004/1.
\end{acknowledgments}
\facilities{SDO/AIA, SDO/HMI, Solar Orbiter EUI, Hinode/EIS, Hinode/XRT.}
\software{SunPy \citep{SunPyCommunity2020}, aiapy \citep{Barnes2020}, SSW/IDL \citep{Freeland1998}, EISPAC (\url{https://github.com/USNavalResearchLaboratory/eispac})}

\bibliography{bibliography}{}

\begin{thebibliography}{}
\expandafter\ifx\csname natexlab\endcsname\relax\def\natexlab#1{#1}\fi
\providecommand{\url}[1]{\href{#1}{#1}}
\providecommand{\dodoi}[1]{doi:~\href{http://doi.org/#1}{\nolinkurl{#1}}}
\providecommand{\doeprint}[1]{\href{http://ascl.net/#1}{\nolinkurl{http://ascl.net/#1}}}
\providecommand{\doarXiv}[1]{\href{https://arxiv.org/abs/#1}{\nolinkurl{https://arxiv.org/abs/#1}}}

\bibitem[{{Abbo} {et~al.}(2016){Abbo}, {Ofman}, {Antiochos}, {Hansteen},
  {Harra}, {Ko}, {Lapenta}, {Li}, {Riley}, {Strachan}, \& et~al.}]{Abbo2016}
{Abbo}, L., {Ofman}, L., {Antiochos}, S.~K., {et~al.} 2016, \ssr, 201, 55,
  \dodoi{10.1007/s11214-016-0264-1}

\bibitem[{{Altschuler} {et~al.}(1977){Altschuler}, {Levine}, {Stix}, \&
  {Harvey}}]{Altschuler1977}
{Altschuler}, M.~D., {Levine}, R.~H., {Stix}, M., \& {Harvey}, J. 1977,
  \solphys, 51, 345, \dodoi{10.1007/BF00216372}

\bibitem[{{Antiochos} {et~al.}(2011){Antiochos}, {Miki{\'c}}, {Titov},
  {Lionello}, \& {Linker}}]{Antiochos2011}
{Antiochos}, S.~K., {Miki{\'c}}, Z., {Titov}, V.~S., {Lionello}, R., \&
  {Linker}, J.~A. 2011, \apj, 731, 112, \dodoi{10.1088/0004-637X/731/2/112}

\bibitem[{{Antolin} {et~al.}(2021){Antolin}, {Pagano}, {Testa}, {Petralia}, \&
  {Reale}}]{Antolin2021}
{Antolin}, P., {Pagano}, P., {Testa}, P., {Petralia}, A., \& {Reale}, F. 2021,
  Nature Astronomy, 5, 54, \dodoi{10.1038/s41550-020-1199-8}

\bibitem[{{Aslanyan} {et~al.}(2022){Aslanyan}, {Pontin}, {Scott}, {Higginson},
  {Wyper}, \& {Antiochos}}]{Aslanyan2022}
{Aslanyan}, V., {Pontin}, D.~I., {Scott}, R.~B., {et~al.} 2022, \apj, 931, 96,
  \dodoi{10.3847/1538-4357/ac69ed}

\bibitem[{{Attrill} {et~al.}(2006){Attrill}, {Nakwacki}, {Harra}, {Van
  Driel-Gesztelyi}, {Mandrini}, {Dasso}, \& {Wang}}]{Attrill2006}
{Attrill}, G., {Nakwacki}, M.~S., {Harra}, L.~K., {et~al.} 2006, \solphys, 238,
  117, \dodoi{10.1007/s11207-006-0167-5}

\bibitem[{{Attrill} {et~al.}(2010){Attrill}, {Harra}, {van Driel-Gesztelyi}, \&
  {Wills-Davey}}]{Attrill2010a}
{Attrill}, G.~D.~R., {Harra}, L.~K., {van Driel-Gesztelyi}, L., \&
  {Wills-Davey}, M.~J. 2010, \solphys, 264, 119,
  \dodoi{10.1007/s11207-010-9558-8}

\bibitem[{{Attrill} {et~al.}(2008){Attrill}, {van Driel-Gesztelyi},
  {D{\'e}moulin}, {Zhukov}, {Steed}, {Harra}, {Mandrini}, \&
  {Linker}}]{Attrill2008}
{Attrill}, G.~D.~R., {van Driel-Gesztelyi}, L., {D{\'e}moulin}, P., {et~al.}
  2008, \solphys, 252, 349, \dodoi{10.1007/s11207-008-9255-z}

\bibitem[{{Baker} {et~al.}(2007){Baker}, {van Driel-Gesztelyi}, \&
  {Attrill}}]{Baker2007}
{Baker}, D., {van Driel-Gesztelyi}, L., \& {Attrill}, G.~D.~R. 2007,
  Astronomische Nachrichten, 328, 773, \dodoi{10.1002/asna.200710787}

\bibitem[{{Baker} {et~al.}(2021){Baker}, {Mihailescu}, {D{\'e}moulin}, {Green},
  {van Driel-Gesztelyi}, {Valori}, {Brooks}, {Long}, \& {Janvier}}]{Baker2021}
{Baker}, D., {Mihailescu}, T., {D{\'e}moulin}, P., {et~al.} 2021, \solphys,
  296, 103, \dodoi{10.1007/s11207-021-01849-7}

\bibitem[{{Barnes} {et~al.}(2020){Barnes}, {Cheung}, {Bobra}, {Boerner},
  {Chintzoglou}, {Leonard}, {Mumford}, {Padmanabhan}, {Shih}, {Shirman}, \&
  et~al.}]{Barnes2020}
{Barnes}, W., {Cheung}, M., {Bobra}, M., {et~al.} 2020, The Journal of Open
  Source Software, 5, 2801, \dodoi{10.21105/joss.02801}

\bibitem[{{Berghmans} {et~al.}(2021){Berghmans}, {Auch{\`e}re}, {Long},
  {Soubri{\'e}}, {Mierla}, {Zhukov}, {Sch{\"u}hle}, {Antolin}, {Harra},
  {Parenti}, \& et~al.}]{Berghmans2021}
{Berghmans}, D., {Auch{\`e}re}, F., {Long}, D.~M., {et~al.} 2021, \aap, 656,
  L4, \dodoi{10.1051/0004-6361/202140380}

\bibitem[{{Bewsher} {et~al.}(2008){Bewsher}, {Harrison}, \&
  {Brown}}]{Bewsher2008}
{Bewsher}, D., {Harrison}, R.~A., \& {Brown}, D.~S. 2008, \aap, 478, 897,
  \dodoi{10.1051/0004-6361:20078615}

\bibitem[{{Carmichael}(1964)}]{Carmichael1964}
{Carmichael}, H. 1964, in NASA Special Publication, Vol.~50, 451

\bibitem[{{Chen} {et~al.}(2021){Chen}, {Przybylski}, {Peter}, {Tian},
  {Auch{\`e}re}, \& {Berghmans}}]{Chen2021}
{Chen}, Y., {Przybylski}, D., {Peter}, H., {et~al.} 2021, \aap, 656, L7,
  \dodoi{10.1051/0004-6361/202140638}

\bibitem[{{Cheng} {et~al.}(2012){Cheng}, {Zhang}, {Saar}, \&
  {Ding}}]{Cheng2012}
{Cheng}, X., {Zhang}, J., {Saar}, S.~H., \& {Ding}, M.~D. 2012, \apj, 761, 62,
  \dodoi{10.1088/0004-637X/761/1/62}

\bibitem[{{Couvidat} {et~al.}(2016){Couvidat}, {Schou}, {Hoeksema}, {Bogart},
  {Bush}, {Duvall}, {Liu}, {Norton}, \& {Scherrer}}]{Couvidat2016}
{Couvidat}, S., {Schou}, J., {Hoeksema}, J.~T., {et~al.} 2016, \solphys, 291,
  1887, \dodoi{10.1007/s11207-016-0957-3}

\bibitem[{{Cranmer}(2009)}]{Cranmer2009}
{Cranmer}, S.~R. 2009, Living Reviews in Solar Physics, 6, 3,
  \dodoi{10.12942/lrsp-2009-3}

\bibitem[{{Cranmer} {et~al.}(2017){Cranmer}, {Gibson}, \&
  {Riley}}]{Cranmer2017}
{Cranmer}, S.~R., {Gibson}, S.~E., \& {Riley}, P. 2017, \ssr, 212, 1345,
  \dodoi{10.1007/s11214-017-0416-y}

\bibitem[{{Culhane} {et~al.}(2007){Culhane}, {Harra}, {James}, {Al-Janabi},
  {Bradley}, {Chaudry}, {Rees}, {Tandy}, {Thomas}, {Whillock}, \&
  et~al.}]{Culhane2007}
{Culhane}, J.~L., {Harra}, L.~K., {James}, A.~M., {et~al.} 2007, \solphys, 243,
  19, \dodoi{10.1007/s01007-007-0293-1}

\bibitem[{{Dissauer} {et~al.}(2019){Dissauer}, {Veronig}, {Temmer}, \&
  {Podladchikova}}]{Dissauer2019}
{Dissauer}, K., {Veronig}, A.~M., {Temmer}, M., \& {Podladchikova}, T. 2019,
  \apj, 874, 123, \dodoi{10.3847/1538-4357/ab0962}

\bibitem[{{Dissauer} {et~al.}(2018{\natexlab{a}}){Dissauer}, {Veronig},
  {Temmer}, {Podladchikova}, \& {Vanninathan}}]{Dissauer2018a}
{Dissauer}, K., {Veronig}, A.~M., {Temmer}, M., {Podladchikova}, T., \&
  {Vanninathan}, K. 2018{\natexlab{a}}, \apj, 863, 169,
  \dodoi{10.3847/1538-4357/aad3c6}

\bibitem[{{Dissauer} {et~al.}(2018{\natexlab{b}}){Dissauer}, {Veronig},
  {Temmer}, {Podladchikova}, \& {Vanninathan}}]{Dissauer2018b}
---. 2018{\natexlab{b}}, \apj, 855, 137, \dodoi{10.3847/1538-4357/aaadb5}

\bibitem[{{Fazakerley} {et~al.}(2016){Fazakerley}, {Harra}, \& {van
  Driel-Gesztelyi}}]{Fazakerley2016}
{Fazakerley}, A.~N., {Harra}, L.~K., \& {van Driel-Gesztelyi}, L. 2016, \apj,
  823, 145, \dodoi{10.3847/0004-637X/823/2/145}

\bibitem[{{Freeland} \& {Handy}(1998)}]{Freeland1998}
{Freeland}, S.~L., \& {Handy}, B.~N. 1998, \solphys, 182, 497,
  \dodoi{10.1023/A:1005038224881}

\bibitem[{{Golub} {et~al.}(2007){Golub}, {DeLuca}, {Austin}, {Bookbinder},
  {Caldwell}, {Cheimets}, {Cirtain}, {Cosmo}, {Reid}, {Sette}, \&
  et~al.}]{Golub2007}
{Golub}, L., {DeLuca}, E., {Austin}, G., {et~al.} 2007, \solphys, 243, 63,
  \dodoi{10.1007/s11207-007-0182-1}

\bibitem[{{Guti{\'e}rrez} {et~al.}(2018){Guti{\'e}rrez}, {Taliashvili}, \&
  {Lazarian}}]{Gutierrez2018}
{Guti{\'e}rrez}, H., {Taliashvili}, L., \& {Lazarian}, A. 2018, \mnras, 479,
  1309, \dodoi{10.1093/mnras/sty1650}

\bibitem[{{Harra} {et~al.}(2015){Harra}, {Baker}, {Edwards}, {Hara}, {Howe}, \&
  {van Driel-Gesztelyi}}]{Harra2015}
{Harra}, L., {Baker}, D., {Edwards}, S.~J., {et~al.} 2015, \solphys, 290, 3203,
  \dodoi{10.1007/s11207-015-0649-4}

\bibitem[{{Harra} {et~al.}(2007){Harra}, {Hara}, {Imada}, {Young}, {Williams},
  {Sterling}, {Korendyke}, \& {Attrill}}]{Harra2007}
{Harra}, L.~K., {Hara}, H., {Imada}, S., {et~al.} 2007, \pasj, 59, S801,
  \dodoi{10.1093/pasj/59.sp3.S801}

\bibitem[{{Harrison} \& {Lyons}(2000)}]{Harrison2000}
{Harrison}, R.~A., \& {Lyons}, M. 2000, \aap, 358, 1097

\bibitem[{{Harvey} \& {Hudson}(1998)}]{Harvey1998}
{Harvey}, K.~L., \& {Hudson}, H.~S. 1998, in Astrophysics and Space Science
  Library, Vol. 229, Observational Plasma Astrophysics : Five Years of YOHKOH
  and Beyond, ed. T.~{Watanabe} \& T.~{Kosugi}, 315,
  \dodoi{10.1007/978-94-011-5220-4_50}

\bibitem[{{Harvey} \& {Recely}(2002)}]{Harvey2002}
{Harvey}, K.~L., \& {Recely}, F. 2002, \solphys, 211, 31,
  \dodoi{10.1023/A:1022469023581}

\bibitem[{{Hassler} {et~al.}(1999){Hassler}, {Dammasch}, {Lemaire}, {Brekke},
  {Curdt}, {Mason}, {Vial}, \& {Wilhelm}}]{Hassler1999}
{Hassler}, D.~M., {Dammasch}, I.~E., {Lemaire}, P., {et~al.} 1999, Science,
  283, 810, \dodoi{10.1126/science.283.5403.810}

\bibitem[{{Heinemann} {et~al.}(2018){Heinemann}, {Temmer}, {Hofmeister},
  {Veronig}, \& {Vennerstr{\o}m}}]{Heinemann2018b}
{Heinemann}, S.~G., {Temmer}, M., {Hofmeister}, S.~J., {Veronig}, A.~M., \&
  {Vennerstr{\o}m}, S. 2018, \apj, 861, 151, \dodoi{10.3847/1538-4357/aac897}

\bibitem[{{Hirayama}(1974)}]{Hirayama1974}
{Hirayama}, T. 1974, \solphys, 34, 323, \dodoi{10.1007/BF00153671}

\bibitem[{{Hofmeister} {et~al.}(2017){Hofmeister}, {Veronig}, {Reiss},
  {Temmer}, {Vennerstrom}, {Vr{\v{s}}nak}, \& {Heber}}]{Hofmeister2017}
{Hofmeister}, S.~J., {Veronig}, A., {Reiss}, M.~A., {et~al.} 2017, \apj, 835,
  268, \dodoi{10.3847/1538-4357/835/2/268}

\bibitem[{{Hudson} {et~al.}(1996){Hudson}, {Acton}, \& {Freeland}}]{Hudson1996}
{Hudson}, H.~S., {Acton}, L.~W., \& {Freeland}, S.~L. 1996, \apj, 470, 629,
  \dodoi{10.1086/177894}

\bibitem[{{Karachik} {et~al.}(2010){Karachik}, {Pevtsov}, \&
  {Abramenko}}]{Karachik2010}
{Karachik}, N.~V., {Pevtsov}, A.~A., \& {Abramenko}, V.~I. 2010, \apj, 714,
  1672, \dodoi{10.1088/0004-637X/714/2/1672}

\bibitem[{{Kong} {et~al.}(2018){Kong}, {Pan}, {Yan}, {Wang}, \&
  {Li}}]{Kong2018}
{Kong}, D.~F., {Pan}, G.~M., {Yan}, X.~L., {Wang}, J.~C., \& {Li}, Q.~L. 2018,
  \apjl, 863, L22, \dodoi{10.3847/2041-8213/aad777}

\bibitem[{{Kopp} \& {Pneuman}(1976)}]{Kopp1976}
{Kopp}, R.~A., \& {Pneuman}, G.~W. 1976, \solphys, 50, 85,
  \dodoi{10.1007/BF00206193}

\bibitem[{{Kosugi} {et~al.}(2007){Kosugi}, {Matsuzaki}, {Sakao}, {Shimizu},
  {Sone}, {Tachikawa}, {Hashimoto}, {Minesugi}, {Ohnishi}, {Yamada}, \&
  et~al.}]{Kosugi2007}
{Kosugi}, T., {Matsuzaki}, K., {Sakao}, T., {et~al.} 2007, \solphys, 243, 3,
  \dodoi{10.1007/s11207-007-9014-6}

\bibitem[{{Kraaikamp} {et~al.}(2023){Kraaikamp}, {Gissot}, {Stegen}, {Mampaey},
  {Verbeeck}, {Auch{\`e}re}, \& {Berghmans}}]{euidatarelease6}
{Kraaikamp}, E., {Gissot}, S., {Stegen}, K., {et~al.} 2023, SolO/EUI Data
  Release 6.0 2023-01, https://doi.org/10.24414/z818-4163

\bibitem[{{Krista} \& {Reinard}(2013)}]{Krista2013}
{Krista}, L.~D., \& {Reinard}, A. 2013, \apj, 762, 91,
  \dodoi{10.1088/0004-637X/762/2/91}

\bibitem[{{Kumar} {et~al.}(2019){Kumar}, {Karpen}, {Antiochos}, {Wyper},
  {DeVore}, \& {DeForest}}]{Kumar2019}
{Kumar}, P., {Karpen}, J.~T., {Antiochos}, S.~K., {et~al.} 2019, \apj, 873, 93,
  \dodoi{10.3847/1538-4357/ab04af}

\bibitem[{{Lemen} {et~al.}(2012){Lemen}, {Title}, {Akin}, {Boerner}, {Chou},
  {Drake}, {Duncan}, {Edwards}, {Friedlaender}, {Heyman}, \&
  et~al.}]{Lemen2012}
{Lemen}, J.~R., {Title}, A.~M., {Akin}, D.~J., {et~al.} 2012, \solphys, 275,
  17, \dodoi{10.1007/s11207-011-9776-8}

\bibitem[{{Levine} {et~al.}(1977){Levine}, {Altschuler}, {Harvey}, \&
  {Jackson}}]{Levine1977}
{Levine}, R.~H., {Altschuler}, M.~D., {Harvey}, J.~W., \& {Jackson}, B.~V.
  1977, \apj, 215, 636, \dodoi{10.1086/155398}

\bibitem[{{Liu} {et~al.}(2007){Liu}, {Lee}, {Yurchyshyn}, {Deng}, {Cho},
  {Karlick{\'y}}, \& {Wang}}]{Liu2007}
{Liu}, C., {Lee}, J., {Yurchyshyn}, V., {et~al.} 2007, \apj, 669, 1372,
  \dodoi{10.1086/521644}

\bibitem[{{Long} {et~al.}(2023){Long}, {Chitta}, {Baker}, {Hannah},
  {Ngampoopun}, {Berghmans}, {Zhukov}, \& {Teriaca}}]{Long2023}
{Long}, D.~M., {Chitta}, L.~P., {Baker}, D., {et~al.} 2023, \apj, 944, 19,
  \dodoi{10.3847/1538-4357/acb0c9}

\bibitem[{{L{\"o}rin{\v{c}}{\'\i}k} {et~al.}(2021){L{\"o}rin{\v{c}}{\'\i}k},
  {Dud{\'\i}k}, {Aulanier}, {Schmieder}, \& {Golub}}]{Lorincik2021}
{L{\"o}rin{\v{c}}{\'\i}k}, J., {Dud{\'\i}k}, J., {Aulanier}, G., {Schmieder},
  B., \& {Golub}, L. 2021, \apj, 906, 62, \dodoi{10.3847/1538-4357/abc8f6}

\bibitem[{{Lowder} {et~al.}(2017){Lowder}, {Qiu}, \& {Leamon}}]{Lowder2017}
{Lowder}, C., {Qiu}, J., \& {Leamon}, R. 2017, \solphys, 292, 18,
  \dodoi{10.1007/s11207-016-1041-8}

\bibitem[{{Mandrini} {et~al.}(2007){Mandrini}, {Nakwacki}, {Attrill}, {van
  Driel-Gesztelyi}, {D{\'e}moulin}, {Dasso}, \& {Elliott}}]{Mandrini2007}
{Mandrini}, C.~H., {Nakwacki}, M.~S., {Attrill}, G., {et~al.} 2007, \solphys,
  244, 25, \dodoi{10.1007/s11207-007-9020-8}

\bibitem[{{McIntosh} {et~al.}(2007){McIntosh}, {Leamon}, {Davey}, \&
  {Wills-Davey}}]{Mcintosh2007}
{McIntosh}, S.~W., {Leamon}, R.~J., {Davey}, A.~R., \& {Wills-Davey}, M.~J.
  2007, \apj, 660, 1653, \dodoi{10.1086/512665}

\bibitem[{{Morgan} \& {Druckm{\"u}ller}(2014)}]{Morgan2014}
{Morgan}, H., \& {Druckm{\"u}ller}, M. 2014, \solphys, 289, 2945,
  \dodoi{10.1007/s11207-014-0523-9}

\bibitem[{{M{\"u}ller} {et~al.}(2020){M{\"u}ller}, {St. Cyr}, {Zouganelis},
  {Gilbert}, {Marsden}, {Nieves-Chinchilla}, {Antonucci}, {Auch{\`e}re},
  {Berghmans}, {Horbury}, \& et~al.}]{Muller2020}
{M{\"u}ller}, D., {St. Cyr}, O.~C., {Zouganelis}, I., {et~al.} 2020, \aap, 642,
  A1, \dodoi{10.1051/0004-6361/202038467}

\bibitem[{{Nitta} \& {Mulligan}(2017)}]{Nitta2017}
{Nitta}, N.~V., \& {Mulligan}, T. 2017, \solphys, 292, 125,
  \dodoi{10.1007/s11207-017-1147-7}

\bibitem[{{Nitta} {et~al.}(2021){Nitta}, {Mulligan}, {Kilpua}, {Lynch},
  {Mierla}, {O'Kane}, {Pagano}, {Palmerio}, {Pomoell}, {Richardson}, \&
  et~al.}]{Nitta2021}
{Nitta}, N.~V., {Mulligan}, T., {Kilpua}, E. K.~J., {et~al.} 2021, \ssr, 217,
  82, \dodoi{10.1007/s11214-021-00857-0}

\bibitem[{{Pariat} {et~al.}(2009){Pariat}, {Antiochos}, \&
  {DeVore}}]{Pariat2009}
{Pariat}, E., {Antiochos}, S.~K., \& {DeVore}, C.~R. 2009, \apj, 691, 61,
  \dodoi{10.1088/0004-637X/691/1/61}

\bibitem[{{Parker}(1983)}]{Parker1983}
{Parker}, E.~N. 1983, \apj, 264, 642, \dodoi{10.1086/160637}

\bibitem[{{Pesnell} {et~al.}(2012){Pesnell}, {Thompson}, \&
  {Chamberlin}}]{Pesnell2012}
{Pesnell}, W.~D., {Thompson}, B.~J., \& {Chamberlin}, P.~C. 2012, \solphys,
  275, 3, \dodoi{10.1007/s11207-011-9841-3}

\bibitem[{{Poletto}(2015)}]{Poletto2015}
{Poletto}, G. 2015, Living Reviews in Solar Physics, 12, 7,
  \dodoi{10.1007/lrsp-2015-7}

\bibitem[{{Raouafi} {et~al.}(2023){Raouafi}, {Stenborg}, {Seaton}, {Wang},
  {Wang}, {DeForest}, {Bale}, {Drake}, {Uritsky}, {Karpen}, \&
  et~al.}]{Raouafi2023}
{Raouafi}, N.~E., {Stenborg}, G., {Seaton}, D.~B., {et~al.} 2023, \apj, 945,
  28, \dodoi{10.3847/1538-4357/acaf6c}

\bibitem[{{Rappazzo} {et~al.}(2012){Rappazzo}, {Matthaeus}, {Ruffolo},
  {Servidio}, \& {Velli}}]{Rappazzo2012}
{Rappazzo}, A.~F., {Matthaeus}, W.~H., {Ruffolo}, D., {Servidio}, S., \&
  {Velli}, M. 2012, \apjl, 758, L14, \dodoi{10.1088/2041-8205/758/1/L14}

\bibitem[{{Reinard} \& {Biesecker}(2008)}]{Reinard2008}
{Reinard}, A.~A., \& {Biesecker}, D.~A. 2008, \apj, 674, 576,
  \dodoi{10.1086/525269}

\bibitem[{{Reiss} {et~al.}(2016){Reiss}, {Temmer}, {Veronig}, {Nikolic},
  {Vennerstrom}, {Sch{\"o}ngassner}, \& {Hofmeister}}]{Reiss2016}
{Reiss}, M.~A., {Temmer}, M., {Veronig}, A.~M., {et~al.} 2016, Space Weather,
  14, 495, \dodoi{10.1002/2016SW001390}

\bibitem[{{Rochus} {et~al.}(2020){Rochus}, {Auch{\`e}re}, {Berghmans}, {Harra},
  {Schmutz}, {Sch{\"u}hle}, {Addison}, {Appourchaux}, {Aznar Cuadrado},
  {Baker}, \& et~al.}]{Rochus2020}
{Rochus}, P., {Auch{\`e}re}, F., {Berghmans}, D., {et~al.} 2020, \aap, 642, A8,
  \dodoi{10.1051/0004-6361/201936663}

\bibitem[{{Rouillard} {et~al.}(2020){Rouillard}, {Pinto}, {Vourlidas}, {De
  Groof}, {Thompson}, {Bemporad}, {Dolei}, {Indurain}, {Buchlin}, {Sasso}, \&
  et~al.}]{Rouillard2020}
{Rouillard}, A.~P., {Pinto}, R.~F., {Vourlidas}, A., {et~al.} 2020, \aap, 642,
  A2, \dodoi{10.1051/0004-6361/201935305}

\bibitem[{{Rust}(1983)}]{Rust1983}
{Rust}, D.~M. 1983, \ssr, 34, 21, \dodoi{10.1007/BF00221193}

\bibitem[{{Scherrer} {et~al.}(2012){Scherrer}, {Schou}, {Bush}, {Kosovichev},
  {Bogart}, {Hoeksema}, {Liu}, {Duvall}, {Zhao}, {Title}, \&
  et~al.}]{Scherrer2012}
{Scherrer}, P.~H., {Schou}, J., {Bush}, R.~I., {et~al.} 2012, \solphys, 275,
  207, \dodoi{10.1007/s11207-011-9834-2}

\bibitem[{{Shibata} {et~al.}(1992){Shibata}, {Ishido}, {Acton}, {Strong},
  {Hirayama}, {Uchida}, {McAllister}, {Matsumoto}, {Tsuneta}, {Shimizu}, \&
  et~al.}]{Shibata1992}
{Shibata}, K., {Ishido}, Y., {Acton}, L.~W., {et~al.} 1992, \pasj, 44, L173

\bibitem[{{Sturrock}(1968)}]{Sturrock1968}
{Sturrock}, P.~A. 1968, in Structure and Development of Solar Active Regions,
  ed. K.~O. {Kiepenheuer}, Vol.~35, 471

\bibitem[{{SunPy Community} {et~al.}(2020){SunPy Community}, {Barnes}, {Bobra},
  {Christe}, {Freij}, {Hayes}, {Ireland}, {Mumford}, {Perez-Suarez}, {Ryan}, \&
  et~al.}]{SunPyCommunity2020}
{SunPy Community}, {Barnes}, W.~T., {Bobra}, M.~G., {et~al.} 2020, \apj, 890,
  68, \dodoi{10.3847/1538-4357/ab4f7a}

\bibitem[{{Tenerani} {et~al.}(2016){Tenerani}, {Velli}, \&
  {DeForest}}]{Tenerani2016}
{Tenerani}, A., {Velli}, M., \& {DeForest}, C. 2016, \apjl, 825, L3,
  \dodoi{10.3847/2041-8205/825/1/L3}

\bibitem[{{Thompson} {et~al.}(2000){Thompson}, {Cliver}, {Nitta},
  {Delann{\'e}e}, \& {Delaboudini{\`e}re}}]{Thompson2000}
{Thompson}, B.~J., {Cliver}, E.~W., {Nitta}, N., {Delann{\'e}e}, C., \&
  {Delaboudini{\`e}re}, J.~P. 2000, \grl, 27, 1431,
  \dodoi{10.1029/1999GL003668}

\bibitem[{{Tian} {et~al.}(2021){Tian}, {Harra}, {Baker}, {Brooks}, \&
  {Xia}}]{Tian2021}
{Tian}, H., {Harra}, L., {Baker}, D., {Brooks}, D.~H., \& {Xia}, L. 2021,
  \solphys, 296, 47, \dodoi{10.1007/s11207-021-01792-7}

\bibitem[{{Tian} {et~al.}(2012){Tian}, {McIntosh}, {Xia}, {He}, \&
  {Wang}}]{Tian2012}
{Tian}, H., {McIntosh}, S.~W., {Xia}, L., {He}, J., \& {Wang}, X. 2012, \apj,
  748, 106, \dodoi{10.1088/0004-637X/748/2/106}

\bibitem[{{Tu} {et~al.}(2005){Tu}, {Zhou}, {Marsch}, {Xia}, {Zhao}, {Wang}, \&
  {Wilhelm}}]{Tu2005}
{Tu}, C.-Y., {Zhou}, C., {Marsch}, E., {et~al.} 2005, Science, 308, 519,
  \dodoi{10.1126/science.1109447}

\bibitem[{{van Driel-Gesztelyi} {et~al.}(2012){van Driel-Gesztelyi}, {Culhane},
  {Baker}, {D{\'e}moulin}, {Mandrini}, {DeRosa}, {Rouillard}, {Opitz},
  {Stenborg}, {Vourlidas}, \& et~al.}]{vanDriel-Gesztelyi2012}
{van Driel-Gesztelyi}, L., {Culhane}, J.~L., {Baker}, D., {et~al.} 2012,
  \solphys, 281, 237, \dodoi{10.1007/s11207-012-0076-8}

\bibitem[{{Vanninathan} {et~al.}(2018){Vanninathan}, {Veronig}, {Dissauer}, \&
  {Temmer}}]{Vanninathan2018}
{Vanninathan}, K., {Veronig}, A.~M., {Dissauer}, K., \& {Temmer}, M. 2018,
  \apj, 857, 62, \dodoi{10.3847/1538-4357/aab09a}

\bibitem[{{Veronig} {et~al.}(2019){Veronig}, {G{\"o}m{\"o}ry}, {Dissauer},
  {Temmer}, \& {Vanninathan}}]{Veronig2019}
{Veronig}, A.~M., {G{\"o}m{\"o}ry}, P., {Dissauer}, K., {Temmer}, M., \&
  {Vanninathan}, K. 2019, \apj, 879, 85, \dodoi{10.3847/1538-4357/ab2712}

\bibitem[{{Viall} \& {Borovsky}(2020)}]{Viall2020}
{Viall}, N.~M., \& {Borovsky}, J.~E. 2020, Journal of Geophysical Research
  (Space Physics), 125, e26005, \dodoi{10.1029/2018JA026005}

\bibitem[{{Wang} {et~al.}(2010){Wang}, {Robbrecht}, {Rouillard}, {Sheeley}, \&
  {Thernisien}}]{Wang2010}
{Wang}, Y.~M., {Robbrecht}, E., {Rouillard}, A.~P., {Sheeley}, N.~R., J., \&
  {Thernisien}, A.~F.~R. 2010, \apj, 715, 39,
  \dodoi{10.1088/0004-637X/715/1/39}

\bibitem[{{Wang} \& {Sheeley}(1990)}]{Wang1990}
{Wang}, Y.~M., \& {Sheeley}, N.~R., J. 1990, \apj, 355, 726,
  \dodoi{10.1086/168805}

\bibitem[{{Wang} \& {Sheeley}(2004)}]{Wang2004}
---. 2004, \apj, 612, 1196, \dodoi{10.1086/422711}

\bibitem[{{Wyper} {et~al.}(2018){Wyper}, {DeVore}, {Karpen}, {Antiochos}, \&
  {Yeates}}]{Wyper2018a}
{Wyper}, P.~F., {DeVore}, C.~R., {Karpen}, J.~T., {Antiochos}, S.~K., \&
  {Yeates}, A.~R. 2018, \apj, 864, 165, \dodoi{10.3847/1538-4357/aad9f7}

\bibitem[{{Yang} {et~al.}(2011){Yang}, {Zhang}, {Li}, \& {Liu}}]{Yang2011}
{Yang}, S., {Zhang}, J., {Li}, T., \& {Liu}, Y. 2011, \apjl, 732, L7,
  \dodoi{10.1088/2041-8205/732/1/L7}

\bibitem[{{Yardley} {et~al.}(2021{\natexlab{a}}){Yardley}, {Brooks}, \&
  {Baker}}]{Yardley2021b}
{Yardley}, S.~L., {Brooks}, D.~H., \& {Baker}, D. 2021{\natexlab{a}}, \aap,
  650, L10, \dodoi{10.1051/0004-6361/202141131}

\bibitem[{{Yardley} {et~al.}(2021{\natexlab{b}}){Yardley}, {Pagano}, {Mackay},
  \& {Upton}}]{Yardley2021a}
{Yardley}, S.~L., {Pagano}, P., {Mackay}, D.~H., \& {Upton}, L.~A.
  2021{\natexlab{b}}, \aap, 652, A160, \dodoi{10.1051/0004-6361/202141142}

\bibitem[{{Yokoyama} \& {Shibata}(1995)}]{Yokoyama1995}
{Yokoyama}, T., \& {Shibata}, K. 1995, \nat, 375, 42, \dodoi{10.1038/375042a0}

\bibitem[{{Young}(2015)}]{Young2015}
{Young}, P.~R. 2015, \apj, 801, 124, \dodoi{10.1088/0004-637X/801/2/124}

\end{thebibliography}
\bibliographystyle{aasjournal}

\end{document}